# The FRET-based structural dynamics challenge - community contributions to consistent and open science practices

Eitan Lerner[1]\*, Benjamin Ambrose[2], Anders Barth[3], Victoria Birkedal[4], Scott C. Blanchard[5], Richard Börner[6], Thorben Cordes[7], Timothy D. Craggs[2], Taekjip Ha[8], Gilad Haran[9], Thorsten Hugel[10], Antonino Ingargiola[11], Achillefs Kapanidis[12], Don C. Lamb[13], Ted Laurence[14], Nam ki Lee[15], Edward A. Lemke[16], Emmanuel Margeat[17], Jens Michaelis[18], Xavier Michalet[11], Daniel Nettels[19], Thomas-Otavio Peulen[20], Benjamin Schuler[19], Claus A.M. Seidel[3], Hamid Soleimaninejad[21], Shimon Weiss[11,22]\*

[1] Department of Biological Chemistry, The Alexander Silberman Institute of Life Sciences, Faculty of Mathematics & Science, The Edmond J. Safra Campus, The Hebrew University of Jerusalem, Jerusalem 9190401, Israel; The Center for Nanoscience and Nanotechnology , The Hebrew University of Jerusalem , Jerusalem 9190401 , Israel

[2] Department of Chemistry, University of Sheffield, Sheffield, UK

[3] Lehrstuhl für Molekulare Physikalische Chemie, Heinrich-Heine-Universität, Düsseldorf, Germany

[4] Department of Chemistry and iNANO center, Aarhus University, 8000 Aarhus, Denmark.

[5] Department of Structural Biology, St. Jude Children's Research Hospital, Memphis, TN, USA

[6] Laserinstitut HS Mittweida, University of Applied Science Mittweida, 09648 Mittweida, Germany

[7] Physical and Synthetic Biology, Faculty of Biology, Ludwig-Maximilians-Universität München, Planegg-Martinsried, Germany

[8] Department of Biophysics and Biophysical Chemistry, Johns Hopkins University School of Medicine, Baltimore, MD 21205, USA; Department of Biophysics and Biophysical Chemistry, Johns Hopkins University School of Medicine, Baltimore, MD 21205, USA; Department of Biomedical Engineering, Johns Hopkins University, Baltimore, MD 21205, USA; Howard Hughes Medical Institute, Baltimore, MD 21205, USA

[9] Department of Chemical and Biological Physics, Weizmann Institute of Science, POB 26, 7610001 Rehovot, Israel

[10] Institute of Physical Chemistry and Signalling Research Centres BIOSS and CIBSS, University of Freiburg, Albertstraße 23a, D-79104 Freiburg, Germany

[11] Department of Chemistry and Biochemistry, Department of Physiology, University of California, Los Angeles, USA

[12] Biological Physics Research Group, Clarendon Laboratory, Department of Physics, University of Oxford, Oxford, United Kingdom

[13] Physical Chemistry, Department of Chemistry, Center for Nanoscience (CeNS), Center for Integrated Protein Science Munich (CIPSM) and Nanosystems Initiative Munich (NIM), Ludwig-Maximilians-Universität, München, 81377, Germany

[14] Physical and Life Sciences Directorate, Lawrence Livermore National Laboratory, Livermore, California

[15] School of Chemistry, Seoul National University, Seoul, South Korea

[16] Departments of Biology and Chemistry, Johannes Gutenberg University Mainz, Mainz, Germany; Institute of Molecular Biology (IMB), Johannes Gutenberg University Mainz, Mainz, Germany; Structural and Computational Biology Unit and Cell Biology and Biophysics Unit, EMBL, Heidelberg, Germany

[17] Centre de Biochimie Structurale (CBS), CNRS, INSERM, Univ Montpellier, 34090 Montpellier, France

[18] Institut of Biophysics, Ulm University, Ulm, Germany.

[19] Department of Biochemistry and Department of Physics, University of Zurich, Zurich, Switzerland

[20] Department of Bioengineering and Therapeutic Sciences, University of California, San Francisco, Mission Bay Byers Hall, 1700 4th Street, San Francisco, CA 94143, USA
[21] Biological Optical Microscopy Platform (BOMP), University of Melbourne, Parkville 3010, VIC, Australia
[22] Department of Physiology, University of California, Los Angeles 90095, USA; California NanoSystems Institute, University of California, Los Angeles 90095, USA; California NanoSystems Institute, University of California, Los Angeles 90095, USA

* corresponding authors: eitan.lerner@mail.huji.ac.il ; sweiss@chem.ucla.edu

## Abstract

Single-molecule Förster resonance energy transfer (smFRET) has become a mainstream technique for probing biomolecular structural dynamics. The rapid and wide adoption of the technique by an ever-increasing number of groups has generated many improvements and variations in the technique itself, in methods for sample preparation and characterization, in analysis of the data from such experiments, and in analysis codes and algorithms. Recently, several labs that employ smFRET have joined forces to try to bring the smFRET community together in adopting a consensus on how to perform experiments and analyze results for achieving quantitative structural information. These recent efforts include multi-lab blind-tests to assess the accuracy and precision of smFRET between different labs using different procedures, the formal assembly of the FRET community and development of smFRET procedures to be considered for entries in the wwPDB. Here we delve into the different approaches and viewpoints in the field. This position paper describes the current "state-of-the field", points to unresolved methodological issues for quantitative structural studies, provides a set of 'soft recommendations' about which an emerging consensus exists, and a list of resources that are openly available. To make further progress, we strongly encourage 'open science' practices. We hope that this position paper will provide a roadmap for newcomers to the field, as well as a reference for seasoned practitioners.

## 1. Introduction

Understanding how biomolecules and their complexes function dynamically is at the heart of several disciplines. Capturing atomic-resolution structural information of the ensemble of states a macromolecular machine populates while it performs its biological function remains an outstanding goal in biology. Linking conformational states to biochemical function requires the ability to resolve precise information on the structure and dynamics of a biological system, which is often altered upon ligand binding or influenced by the chemical and physical properties of its environment. Conventional structural tools provide detailed 'snapshots' of states in a crystallized or frozen equilibrium (e.g. X-ray crystallography and single-particle cryo-EM) or an ensemble average of all contributing conformations (e.g. NMR, SAXS or SANS). However, steady-state structural ensembles of interconverting conformational states, or structures of reaction intermediates, and in particular of short-lived intermediates (e.g. along a reaction pathway), are hard to capture and characterize with classical structural biology techniques.

Various methods for studying macromolecular dynamics have been introduced over the years. For example, NMR(Palmer, 2004; Clore and Iwahara, 2009; Ravera *et al.*, 2014; Anthis and Clore, 2015) and EPR(Krstić *et al.*, 2011; Jeschke, 2012, 2018) techniques to study conformational dynamics and capture transient intermediates have been applied broadly. Time-resolved crystallographic investigations on biological macromolecules, which use complex trapping and time-resolved approaches (pump-probe, ultra-fast mixing), are able to resolve functionally-relevant structural displacements in reaction intermediates(Schlichting *et al.*, 1990; Schlichting and Chu, 2000; Moffat, 2001; Schotte *et al.*, 2003; Kupitz *et al.*, 2014). Recent advances in developing microfluidic mixing/spraying devices for time-resolved cryoEM(Feng *et al.*, 2017; Kaledhonkar *et al.*, 2018), cross-linking mass spectrometry (XL-MS), as well as progress in computational methods (Murakami *et al.*, 2013; Slavin and Kalisman, 2018; Braitbard, Schneidman-Duhovny and Kalisman, 2019; Brodie *et al.*, 2019; Iacobucci *et al.*, 2019), provide novel tools for dynamic structure determination and indicate the growing importance of methods able to directly and continuously capture dynamical structures.

Förster resonance energy transfer (FRET; frequently also referred to as fluorescence resonance energy transfer) studies at the ensemble(Grinvald, Haas and Steinberg, 1972; Haas *et al.*, 1975; Haas and Steinberg, 1984; Hochstrasser, Chen and Millar, 1992; Peulen, Opanasyuk and Seidel, 2017) and single-molecule(Ha *et al.*, 1996; Deniz *et al.*, 1999; Lerner *et al.*, 2018) levels have emerged as an important tool to map the heterogeneity of biomolecules in this era of "*dynamic structural biology*", as well as to obtain information on structural dynamics (both fluctuations within conformational states and transitions between different conformational states) over time scales ranging from nanoseconds to minutes(Schuler and Hofmann, 2013; Mazal and Haran, 2019). FRET has been used extensively to study conformational changes under steady-state conditions(Schuler *et al.*, 2005; Lerner *et al.*, 2018) as well as their dynamics(Zhuang *et al.*, 2000; Schuler, Lipman and Eaton, 2002; Lipman *et al.*, 2003; Margittai *et al.*, 2003). FRET studies using fluorescence lifetime measurement techniques at the ensemble level(Grinvald, Haas and Steinberg, 1972; Haas *et al.*, 1975; Haas and Steinberg, 1984; Hochstrasser, Chen and Millar, 1992; Peulen, Opanasyuk and Seidel, 2017) have long been used to analyze the heterogeneities in molecular systems that exist on time-scales longer than the lifetime of the fluorophore (typically longer than a few ns). These nanosecond time resolution techniques have also been utilized to reveal rapid structural dynamics between sub-states in populations of single-molecules(Neubauer *et al.*, 2007; Gansen *et al.*, 2018). Beyond traditional ensemble FRET, applications of FRET between inorganic probes that are brighter or have long fluorescence lifetimes, such as nanoparticles and lanthanides, have recently been discussed(Guo *et al.*, 2019; Léger *et al.*, 2020). Single molecule FRET (smFRET; also known as single-pair FRET or spFRET) has given otherwise unattainable information into biomolecular conformational dynamics and biomolecular interactions(Michalet, Weiss and Jäger, 2006; Sasmal *et al.*, 2016; Lerner *et al.*, 2018; Mazal and Haran, 2019).

Many techniques can determine structural ensembles. More recently, distances derived from smFRET experiments have been utilized as spatial restraints to computationally determine potential structural models(Margittai *et al.*, 2003; Woźniak *et al.*, 2008; Muschielok *et al.*, 2008; Sindbert *et al.*, 2011; Craggs and Kapanidis, 2012; Kalinin *et al.*, 2012; McCann *et al.*, 2012; Dimura *et al.*, 2016; Hellenkamp *et al.*, 2017; Lerner, Ingargiola and Weiss, 2018; Craggs *et al.*, 2019; Schuler *et al.*, 2020) and conformational ensembles(Borgia *et al.*, 2018; Holmstrom *et al.*, 2018). As an example, we show the outcome of a multimodal smFRET study on the conformational landscape of a 12-mer chromatin array in Fig 1B (Kilic *et al.*, 2018), with a dynamics occurring at timescales from nanoseconds to >100 seconds. A unique aspect of smFRET is that structural, kinetic and spectroscopic information can be simultaneously recorded in a single experiment. This facilitates the linking of dynamic and structural information in an integrative approach (Fig. 1A), and potentially reduces the space of structures in dynamic exchange that are compatible with the experimental data to resolve structures of dynamic states and distinguish between competing structural models(Hellenkamp *et al.*, 2017; Kilic *et al.*, 2018; Yanez Orozco *et al.*, 2018; Sanabria *et al.*, 2020) (see Fig. 1). This type of approach to visualizing biomolecules in action under ambient conditions has emphasized the importance of the dynamical aspect of their action, resolving transitions between various conformational states, which are the physical basis for their function (Henzler-Wildman *et al.*, 2007; Aviram *et al.*, 2018; Lerner, Ingargiola and Weiss, 2018; Sanabria *et al.*, 2020). The dynamic view beyond the fascinating well-ordered static structures of chromatin fibers (Fig. 1B, top panel) reveals their dynamic structural heterogeneity (Fig. 1B, bottom), which is essential for gene function. Combining the structural features and exchange dynamics of FRET species for all six FRET pairs (Fig. 1B, middle panel), the authors detected that actually >70% of the chromatin adopts half-open or open conformations under physiological conditions. These flexible conformations represent the central interconversion hub for the distinct stacking registers of chromatin that is difficult to detect with conventional structural techniques.

smFRET studies are widely used by hundreds of laboratories worldwide. Measurements are obtained using the two most common formats – surface-immobilized and freely diffusing molecules – and have been acquired and analyzed using mostly custom-built microscopes, various acquisition and analysis software, and oftentimes lab-specific protocols, with data collected and stored in a variety of file formats. A recent multi-laboratory study has shown that reproducible quantitative FRET values, reflecting distances between FRET dyes, can be obtained across different experimental acquisition and data analysis procedures(Hellenkamp *et al.*, 2018).

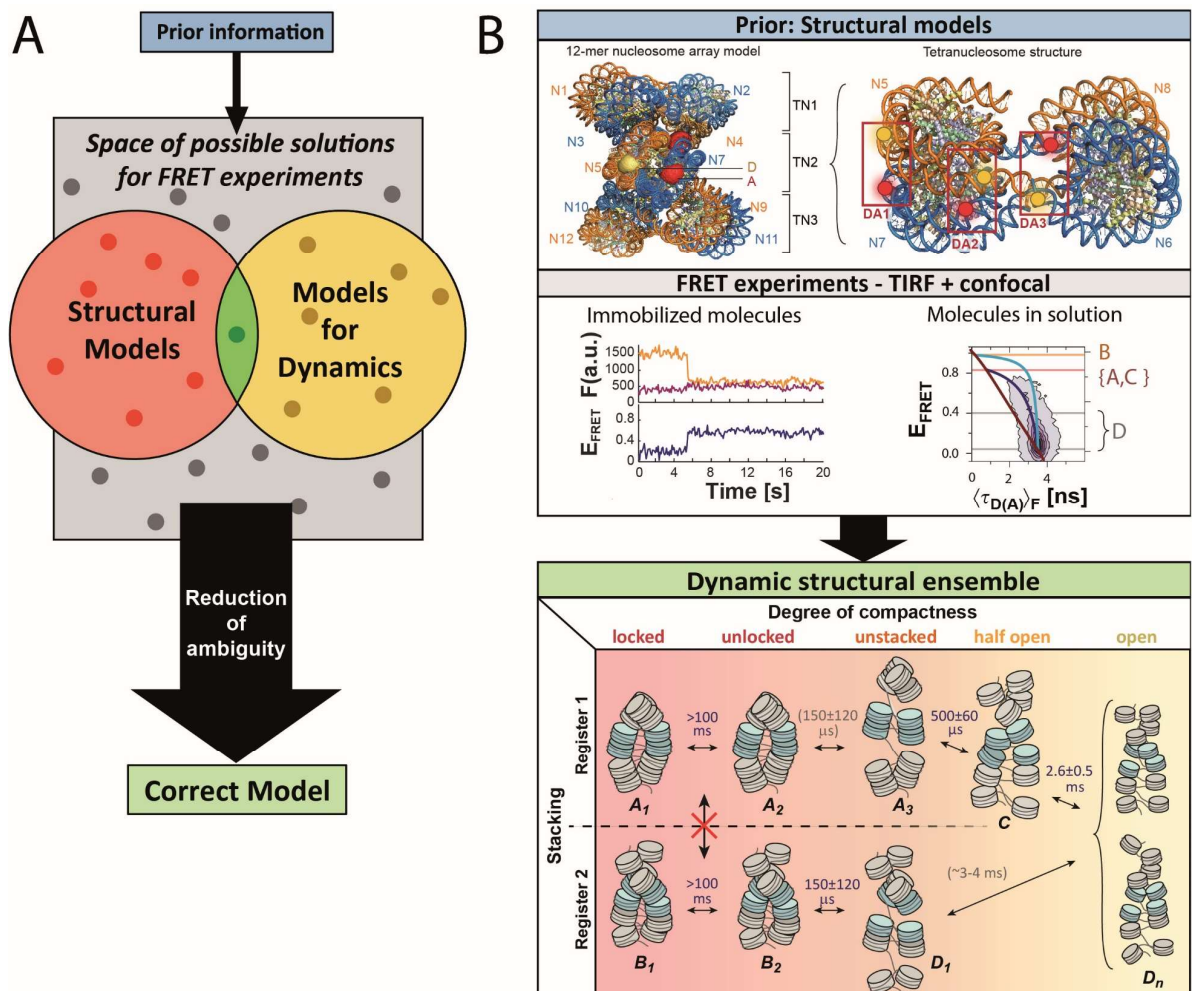

**Fig. 1: (A)** Integrative modeling requires structural *and* dynamic information. Prior information from conventional approaches (X-ray, NMR, cryoEM) together with computational tools defines the space of possible solutions for FRET-assisted structural modeling. The combination of structural (inter-dye distances) and dynamic (kinetic connectivity and exchange rates) information enables identification of a consistent model. **(B)** Study of structure and dynamics of chromatin fibers. Multimodal FRET-study of structure and dynamics of chromatin fibers using three FRET labeling positions (DA1-3) for two pairs of dyes with distinct Förster distances (Eq. 3 bellow). Prior structural information provided by X-ray crystallography of tetra-nucleosome(top, right PDB ID: 1ZBB(Schalch *et al.*, 2005)) and cryo-electron microscopy (top, left(Song *et al.*, 2014)) is combined with the structural and dynamic information obtained by FRET experiments on immobilized molecules measured by total internal reflection (TIRF) microscopy and on freely diffusing molecules by confocal microscopy (Kilic *et al.*, 2018). From the combined information, a consistent model is derived for chromatin fiber conformations with shifted registers, which are connected by slow (> 100 ms) and fast de-compaction processes (150 μs) that do not proceed directly, but rather through an open fiber conformation (Fig. 1B is adapted from figures in ref. (Kilic *et al.*, 2018) with permission).

An important step would be to ensure that smFRET measurements as well as smFRET-derived distances are highly reproducible by different groups. This directly translates into general methodological recommendations, ranging from sample preparation and characterization to setup description, data acquisition and preservation, up to data analysis. An immediate benefit would be a reliable way to validate results and estimate the accuracy and precision of measurements. Here, several laboratories with expertise in smFRET (and ensemble FRET), without pretension to be exhaustive or exclusive, have united to endorse these grass-root efforts and to propose additional steps to organize the community around consistent and open science practices. These recommendations on how to "practice" smFRET should not be viewed as an attempt to regiment the community. On the contrary, our proposal, detailed below, aims at kick-starting a process and an open dialog about existing practices in our field. We believe this proposal will enable the preservation of existing data, data formats and analysis methods, while encouraging and facilitating the innovation that has characterized this field for the past two decades.

## 2. State of the art and limitations of smFRET

smFRET experiments combine the advantages of single-molecule detection, for instance:
- resolving structural and dynamic heterogeneity
- allowing high quality measurements with low concentrations of the molecules of interest, as the sample is analyzed one molecule at a time,
- measuring kinetics without the need to synchronize the sample (such as in stopped-flow experiments),

with the intrinsic ability of FRET to probe distances on the macromolecular scale (~2.5 to 10 nm)(Stryer and Haugland, 1967). Currently, the application of smFRET is expanding rapidly due to its unique capabilities and high sensitivity. There are several areas where new smFRET-based methods are being developed and applied. We highlight a few directions where smFRET will have high impact and discuss the current limitations.

### 2.1 Dynamics

smFRET measurements excel in providing unique insights into the detection and quantification of conformational dynamics. Here we define conformational dynamics as transitions between conformational states (defined by activation barriers larger than some value, usually > $1k_BT$) or fluctuations within states (defined by potential wells between activation barriers). Combining several smFRET experimental modalities, it is possible to detect equilibrium and non-equilibrium dynamics on timescales across twelve orders of magnitude (nanoseconds to thousands of seconds). For dynamics on the order of 10 ms or slower, transitions between conformational states can be directly observed using immobilization-based smFRET approaches, as have been demonstrated in numerous studies(Zhuang *et al.*, 2000; Juette *et al.*, 2014; Deniz, 2016; Sasmal *et al.*, 2016). Using analysis methods such as hidden Markov modeling approaches(Andrec, Levy and Talaga, 2003; McKinney, Joo and Ha, 2006; Munro *et al.*, 2007; Zarrabi *et al.*, 2018; Steffen *et al.*, 2020), the number and connectivity of the states, and the individual transition rates can be extracted from the data. However, it is certainly possible that an identified sub-population may represent rapidly interconverting states, generating a single time-averaged FRET efficiency, for example arising from protein domain movements that can occur on timescales faster than 10 ms(Henzler-Wildman and Kern, 2007). New technologies are emerging to push the time scale of the dynamics that can be resolved directly from smFRET measurements (without analyses of photon statistics) into the sub-millisecond regime (e.g. Metal Enhanced Fluorescence with FRET in plasmonic hotspots(Acuna *et al.*, 2012) or sub-millisecond camera-based imaging(Farooq and Hohlbein, 2015; Fitzgerald *et al.*, 2019)). Of note, these types of approaches have also been shown to have the capability to extend the range to distances observable by smFRET(Baibakov *et al.*, 2019).

Dynamics at the sub-millisecond timescale and even faster can also be retrieved from smFRET data by analyzing (time-correlated) single photon counting statistics, typically on freely diffusing molecules, detected as they diffuse through a confocal observation volume. Several approaches have been developed, including burst variance analysis (BVA)(Torella *et al.*, 2011), two-channel kernel-based density distribution estimator (2CDE)(Tomov *et al.*, 2012), photon distribution analysis, also known as probability distribution analysis (PDA)(Antonik *et al.*, 2006; Nir *et al.*, 2006; Kalinin *et al.*, 2007, 2008, 2010; Santoso, Torella and Kapanidis, 2010), 2D histograms of burst donor mean fluorescence lifetime versus FRET efficiency (Rothwell *et al.*, 2003; Sisamakis *et al.*, 2010), burst recurrence analysis(Hoffmann *et al.*, 2011) and maximum likelihood approaches(Köllner and Wolfrum, 1992; Zander *et al.*, 1996; Maus *et al.*, 2001; Nettels *et al.*, 2007; Chung *et al.*, 2012), such as photon-by-photon hidden Markov modelling(Keller *et al.*, 2014; Pirchi *et al.*, 2016) and recoloring(Gopich and Szabo, 2009; Gopich, 2012; Lerner, Ingargiola and Weiss, 2018). By combining smFRET with fluorescence correlation spectroscopy (FCS)(Magde, Elson and Webb, 1972; Rigler *et al.*, 1993; Widengren *et al.*, 2001; Torres and Levitus, 2007; Gurunathan and Levitus, 2010; Felekyan *et al.*, 2013; Schuler, 2018), it is possible to quantify FRET dynamics faster than the timescale of translational diffusion through the observation volume (i.e. as fast as a few tens of nanoseconds). FRET dynamics as fast as a few picoseconds can also be retrieved from a variant of FCS dubbed '*nanosecond FCS*' (nsFCS) (Nettels, Hoffmann and Schuler, 2008; Schuler and Hofmann, 2013). With the capability to distinguish between different species in the measurements, filtered-FCS(Böhmer *et al.*, 2002; Enderlein *et al.*, 2005; Kapusta *et al.*, 2007; Laurence *et al.*, 2007, 2008; McCann *et al.*, 2012; Felekyan *et al.*, 2013) is a powerful method for extracting the rates of conformational transitions between the different species or sub-populations and the hydrodynamic radii of these species in one go. By that, filtered-FCS coupled to FRET may assist in linking between conformational states and the status of their binding to other biomolecules, hence linking structure to function. In experiments on freely diffusing single molecules, the dynamic processes that can be studied are limited by the diffusion time (with the exception of burst recurrence analysis(Hoffmann *et al.*, 2011)). Thus, to fully cover both fast and slow dynamic processes, experiments on immobilized molecules for which data were registered with slower cameras and experiments on freely-diffusing molecules for which data were registered by fast point detectors are ideally combined (see Figure 1B).

In addition to the dynamics and transitions between conformations, the flexibility within a given conformational state can be studied. The flexibility of a conformation is expressed in its distance distribution and in the rate at which the reported inter-dye distance fluctuates within that distribution. Both considerations are important features of biomolecular conformational dynamics (for an example see Fig. 2). For instance, this information can potentially distinguish between rigid and flexible conformational states. Information regarding the flexibility of a given conformation can be retrieved from the analyses of conformation-related sub-ensemble fluorescence decays(Neubauer *et al.*, 2007; Sisamakis *et al.*, 2010; Lerner *et al.*, 2014; Rahamim *et al.*, 2015; Gansen *et al.*, 2018). It is, however, important to mention that the distinction between dynamics within a conformational state and dynamics of transitions between different conformational states is still under debate, and highly depends on the definition of an activation barrier for different modes of structural dynamics and in the different smFRET modalities.

These examples are just the tip of the iceberg. New approaches are currently being developed that aim to improve the quantification of FRET dynamics information as well as provide a means to better distinguish between complex models. Overall, conformational dynamics is an area in which we foresee smFRET providing significant insights in fields of biology that can otherwise be difficult to obtain using other methods. The ultimate goal is to combine the structural and dynamic information in order to reduce the space of potential solutions for the underlying structures of conformational states (Fig. 1) and to gain detailed information on kinetic pathways between the associated states.

## 2.2 Structural Studies

The use of smFRET for structural determination has recently emerged(Muschielok *et al.*, 2008; Brunger *et al.*, 2011; Treutlein *et al.*, 2012; Nagy *et al.*, 2015; Hellenkamp *et al.*, 2017; Holmstrom *et al.*, 2018; Kilic *et al.*, 2018; Yanez Orozco *et al.*, 2018; Craggs *et al.*, 2019; Sanabria *et al.*, 2020). FRET-based approaches are particularly powerful to study structures of large heterogeneous, flexible and dynamic biomolecules and complexes. Structural characterization using smFRET-derived distance restraints requires: i) preparing and measuring multiple donor-acceptor labeled variants with multiple pairs of labeling positions, each of which will be considered as one reaction coordinate, ii) a large number of control experiments (e.g. activity after labeling or immobilization, photophysics of the probes, dye rotational freedom) and iii) non-trivial transformations from proximity ratios (uncorrected FRET efficiency values), to corrected FRET efficiencies (corrected for donor fluorescence leakage to the acceptor channel, the fraction of acceptor photons following its direct excitation at the wavelength of the donor (and not excitation via energy transfer) and the imbalance in the donor and acceptor fluorescence quantum yields, $\Phi_F$, and in their detection efficiencies), to inter-dye distance information (or equilibrium distance distribution), and inter-residue distance information (Lerner *et al.*, 2018). Factors such as the orientational flexibility of the dyes, changes of $\Phi_F$ due to effects of the immediate local environment in the vicinity of the labeling positions(Steffen, Sigel and Börner, 2016), and uncertainties in the refractive index between the dyes, further complicate accurate distance extraction from the results of smFRET measurements. While complications of this nature can be addressed by a variety of methods, smFRET studies with multiparameter fluorescence detection (MFD)(Rothwell *et al.*, 2003; Sisamakis *et al.*, 2010) nano-second Alternating Laser Excitation (nsALEX)(Laurence *et al.*, 2005) and pulsed-interleaved excitation (PIE)(Müller *et al.*, 2005;

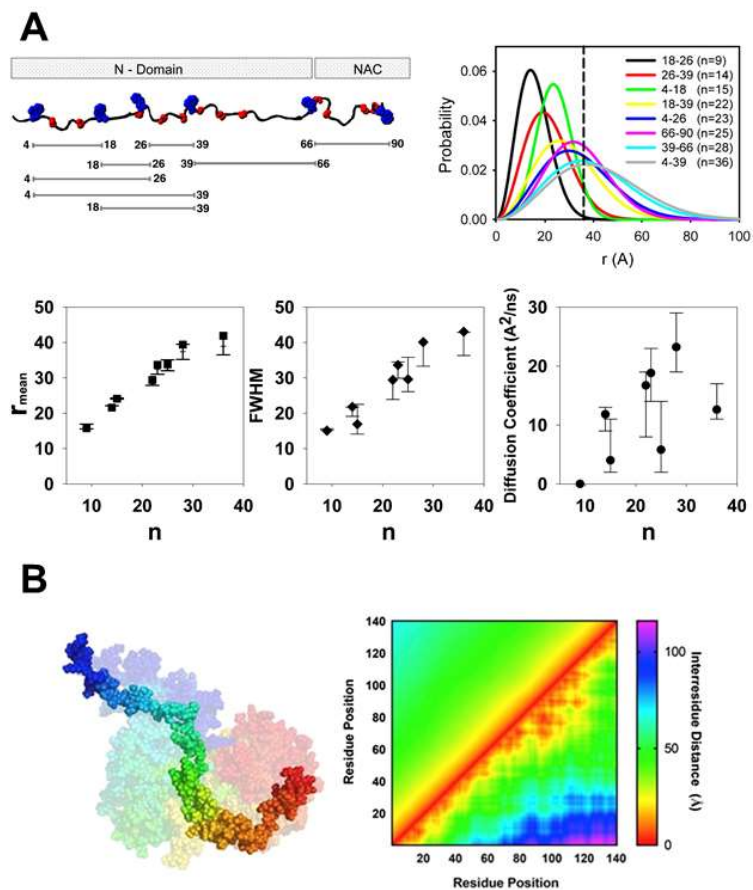

**Fig. 2:** An example for deriving the ensemble structure and flexibility of α-Synuclein, a highly flexible protein, from multiple time-resolved FRET measurements. A – Distance distributions are retrieved from multiple FRET measurements, with the inter-dye diffusion coefficient(Grupi and Haas, 2011). B – Distance distributions from multiple FRET measurements are used to constrain modeling of the ensemble structure(Ferrie *et al.*, 2018). Similar analyses of time-resolved fluorescence data, per sub-population in smFRET measurements, could also be utilized, to retrieve the distance distribution and its flexibility, per each conformational state, as in (Ferrie *et al.*, 2018). (Fig. 2A and 2B are adapted from (Grupi and Haas, 2011) and (Ferrie *et al.*, 2018), respectively, with permission).

Kudryavtsev *et al.*, 2012) (nsALEX and PIE are different implementations of the same technique) were developed to aid in these endeavors by allowing the simultaneous monitoring of fluorescence lifetimes, brightnesses and anisotropies, local dye flexibility and inter-dye distance distributions in fast exchange. The underlying information on the distance between the donor- and acceptor-labeled residues, to be recovered and used in integrative modelling, can also be extracted from time-resolved fluorescence spectroscopy measurements(Grupi and Haas, 2011; Kudryavtsev *et al.*, 2012; Orevi *et al.*, 2014; Dimura *et al.*, 2016; Ferrie *et al.*, 2018). For structural modeling, proper error analysis has to be used throughout(Muschielok *et al.*, 2008; Muschielok and Michaelis, 2011; Kalinin *et al.*, 2012; Hellenkamp *et al.*, 2017). However, a position paper on how to transform the raw time-resolved data to accurate distance information has not yet been published. While analysis methods, appropriate models and the discussion of potential shortcomings of FRET applications to structural studies have been well-established by various groups in the field(Grinvald, Haas and Steinberg, 1972; Hochstrasser, Chen and Millar, 1992; Muschielok *et al.*, 2008; Hellenkamp *et al.*, 2017; Holmstrom *et al.*, 2018; Ingargiola, Weiss and Lerner, 2018), the mechanisms for disseminating and accessing the raw data and techniques employed for time-resolved ensemble, sub-ensemble and single-molecule measurements requires further improvement (e.g. to meet requirements of the wwPDB). The abovementioned developments, as well as others, are expected to assist in studying conformational dynamics of biomolecules. Section 3 will discuss these in more detail.

## 2.3 *Hybrid Methods for Imaging*

SmFRET can also make a big impact when combined with other approaches. For example, the combination of smFRET with data from stimulated emission depletion (STED) microscopy(Hell and Wichmann, 1994; Klar *et al.*, 2000) provides more detailed 3D information(Kim *et al.*, 2018; Günther *et al.*, 2019; Tardif *et al.*, 2019). The combination of fluorescence imaging with spectroscopy makes it possible to detect more species within a pixel of an image, expanding the information that can be extracted from such an experiment. Correlative imaging with electron-microscopy, fluorescence and smFRET(Schirra Jr and Zhang, 2014) also has the potential to allow the recognition of different subpopulations in the sample, which can then be separated for particle reconstructions(Collinson and Verkade, 2019).

## 2.4 *In cell smFRET*

Several groups have shown that smFRET can be performed in live bacterial and eukaryotic cells by using *in vitro* labeled biomolecules that can be internalized in the cells by several means, including electroporation and microinjection. Electroporation relies on transient formation of pores in the cell membrane, which allow labeled biomolecules to enter the cell and get trapped inside as the pores close; cell washing is required prior to imaging to remove any non-internalized molecules, and the efficiency of loading is tunable due to its dependence on the applied voltage. Electroporation has been shown to work well for the internalization of ssDNA, dsDNA, and proteins into bacteria and yeast(Plochowietz, Crawford and Kapanidis, 2014; Sustarsic *et al.*, 2014; A Plochowietz *et al.*, 2016; Craggs *et al.*, 2019). Further work has shown that tRNA is also internalized efficiently and can be used to study the cycle of tRNA utilization before and during RNA translation(Anne Plochowietz *et al.*, 2016; Volkov *et al.*, 2018). The approach is amenable to protein internalization, and both the structure and activity of internalized proteins can be preserved.

Further progress in this field will depend on avoiding the need for electroporation by using cell-permeable dyes for labeling unnatural amino acids with fluorophores that can serve as complementary probes for FRET(Sustarsic and Kapanidis, 2015). Furthermore, it is highly desirable to find ways to synchronize *in vivo* chemical reactions, since this would allow fast reactions to be observed. The use of caging groups and chemical agents that can affect gene expression can help in this regard. Finally, as with many experiments in small cells such as bacteria, a way to control the number of FRET pairs and to slow down photobleaching will help in the collection of large statistical data sets with long observation times, and hence a way to probe many minute-timescale processes in their natural environment(Hell and Wichmann, 1994; Klar *et al.*, 2000; Kim *et al.*, 2018). Microinjection of labeled molecules is an alternative approach, especially for smFRET in live eukaryotic cells, and has been demonstrated to yield structural information and dynamics from nanoseconds to milliseconds.(Sakon and Weninger, 2010; König *et al.*, 2015).

## 2.5 *Combining smFRET with other fluorescence methods*

Several groups have combined smFRET with protein induced fluorescence enhancement (PIFE)(Hwang, Kim and Myong, 2011; Hwang and Myong, 2014; Lerner *et al.*, 2016; Ploetz *et al.*, 2016), photo-induced electron transfer (PET)(Haenni *et al.*, 2013), quenchable FRET(Cordes *et al.*, 2010) and stacking-induced fluorescence increase (SIFI)(Morten, Steinmark and Magennis, 2020) to expand the ruler's distance dynamic range (mostly towards shorter distances). The advantages of combining smFRET with other fluorescence-based rulers is obvious – gaining more spatial information on biomolecular systems being measured, as well as information on possible synchronizations between the systems' different parts and between different modes of motion. For example, using PIFE combined with smFRET, it was possible to correlate certain conformations of the DNA bubble during transcription initiation as associated with the extrusion of pro-motor nontemplate bases outside the transcription complex(Ploetz *et al.*, 2016). Advancements of these hybrid experimental approaches will introduce better data analysis schemes as well as a broader variety of suitable dyes.

## 2.6 *Combining smFRET with nanomanipulation methods*

Several groups have combined smFRET with various manipulation methods, including optical tweezers(Hohng *et al.*, 2007), magnetic tweezers(Swoboda *et al.*, 2014; Long, Parks and Stone, 2014), tethered particle motion - TPM (to introduce the related methods of tethered fluor-ophore motion, TFM (May *et al.*, 2014)), anti-Brownian electrokinetic (ABEL) trap(Wilson and Wang, 2019) and force spectroscopy by DNA origami(Nickels *et al.*, 2016). Limitations of these approaches include the limited resolution of TFM with regard to DNA translocation (as opposed to optical and magnetic tweezer methods). On the other hand, photobleaching is increased in optical tweezer experiments (due to the use of a high power IR laser), and the presence of the beads used for the nanomanipulation creates significant fluorescence background. Further progress in these combined manipulation methods will require high photon counts (combined with means to delay photobleaching) to increase the TFM resolution with translocation on DNA.

## 2.7 *smFRET between multiple chromophores*

For structural studies, the limitations of the single-distance readout provided by single-pair FRET is overcome by measuring many distances in separate experiments(McCann *et al.*, 2012; Dimura *et al.*, 2016; Lerner, Ingargiola and Weiss, 2018). If the biomolecule is found in multiple conformational states, however, then it is challenging to assign the observed species in smFRET to the structural states or to detect coordinated conformational changes unambiguously. By measuring the transfer of excitation energy between three or more spectrally different fluorophores, multiple distances are obtained simultaneously and the correlation of the distances is revealed. Following early ensemble implementations(Horsey *et al.*, 2000; Ramirez-Carrozzi and Kerppola, 2001; Haustein, Jahnz and Schwille, 2003; Watrob, Pan and Barkley, 2003; Yim *et al.*, 2012), three- and four-color FRET experiments have been applied to various static(Clamme and Deniz, 2005; Lee *et al.*, 2007; Stein, Steinhauer and Tinnefeld, 2011) and dynamic systems(Hohng, Joo and Ha, 2004; Lee *et al.*, 2010; Lee, Lee and Hohng, 2010; Ratzke, Hellenkamp and Hugel, 2014; Götz *et al.*, 2016; Wasserman *et al.*, 2016; Vušurović *et al.*, 2017; Morse *et al.*, 2020) at the single molecule level. FRET to many acceptors has also been reported(Uphoff *et al.*, 2010). Multi-color FRET experiments, however, remain challenging due to the higher spectral overlap between the different fluorophores and the use of UV or NIR dyes with less optimal fluorescence properties.

Despite advances in the development of orthogonal labeling approaches(Milles, Koehler, *et al.*, 2012; Milles, Tyagi, *et al.*, 2012), it also remains challenging to attach multiple fluorophores to specific sites in proteins. Especially for diffusion-based experiments, the increased shot noise, higher correction factors and more complex FRET efficiency calculations have limited a more widespread application of multi-color FRET approaches. Recent advances include the development of a photon distribution analysis for three-color FRET to extract three-dimensional distance distributions(Barth, Voith von Voithenberg and Lamb, 2019) and a maximum likelihood approach applied to the study of fast protein folding(Yoo *et al.*, 2018). Further progress in multiple chromophores smFRET will require expanding the ability to work in the near infra-red (which requires better fluorophores and detectors in that region) and ways to use analysis of single-molecule spectra(Lacoste *et al.*, 2000), which may be more efficient than splitting the fluorescence into 3 or 4 individual channels.

*2.8 smFRET with nanomaterials*

Recently, emerging structurally synthesized and target specific nanomaterials such as Quantum dots (QDs)(Jamieson *et al.*, 2007) and Aggregation Induced Emission (AIE) nanoparticles(Hong, Lam and Tang, 2011) have given the opportunity to implement chemically engineered fluorophores with wide applications in structural biology investigations and specifically, in FRET-related applications(Medintz *et al.*, 2003; Oh *et al.*, 2005; Shi *et al.*, 2006; Soleimaninejad *et al.*, 2017). On the other hand, smFRET measurements of these synthetized materials are also challenging(Soleimaninejad *et al.*, 2017). The adjustable and nano-environment specific fluorescence emissions of AIE nanoparticles (NPs)(Hong, Lam and Tang, 2011) make it arduous to calculate the coefficients for these dyes required in smFRET measurements. smFRET investigations with synesthetic QDs and AIE NPs are also particular difficult to decipher due to complex photophysics and electrodynamics properties. Clear smFRET exploration of these distinct nanomaterials are not comprehensively reviewed and available in literature, although their applications are exponentially growing in many fields.

## 3. Outstanding challenges of smFRET as a quantitative structural tool

When using smFRET experiments for structural studies, many steps need to be taken to convert the raw data (photons detected and registered by the detectors) to absolute inter-dye distance measures. A recent community-wide round-robin test, outlined below, indicates how crucial these steps are to obtain comparable results in different laboratories. As a community, we are working on defining viable procedures for extracting absolute FRET efficiencies correctly and thereby the ability to yield precise and accurate inter-dye distance information. Some of the future challenges include:

*3.1 Determination of the Förster Distance*

In FRET, the excitation energy of the donor fluorophore is transferred to an acceptor fluorophore via dipolar coupling. Considering a single donor-acceptor distance, $R_{DA}$, the efficiency, $E$, of this nonradiative transfer process scales with the sixth power of $R_{DA}$ normalized by the Förster distance, $R_0$ (Eq. 1):

$$E = \frac{1}{1+\left(\frac{R_{DA}}{R_0}\right)^6} \qquad \text{(Eq. 1)}$$

However, in smFRET studies, dyes are usually coupled to the biomolecules via long flexible linkers, which results in an equilibrium distribution of $R_{DA}$ values, $p(R_{DA})$. In this case, one may observe a mean FRET efficiency $\langle E \rangle$ related to the mean FRET, averaged over all distance probabilities (Eq. 2).

$$\langle E \rangle = \int_0^\infty \frac{p(R_{DA})}{1+\left(\frac{R_{DA}}{R_0}\right)^6} \, dR_{DA} \qquad \text{(Eq. 2)}$$

Alternatively, one may observe a mean FRET efficiency $\langle E \rangle$ related to a FRET-averaged apparent donor acceptor distance $\langle R_{DA} \rangle_E$ in Eq. 1(Kalinin *et al.*, 2012).

$R_0$ (Eq. 1), the $R_{DA}$ at which 50% of the donor excitation energy is transferred to the acceptor fluorophore. $R_0$ depends on other parameters, including the donor fluorescence $\Phi_{F,D}$, the overlap between the normalized donor emission spectrum $\tilde{F}_D(\lambda)$, and the acceptor excitation spectrum with extinction coefficient $\varepsilon_A(\lambda)$, the relative orientation of the dye dipoles captured by the orientation factor $\kappa^2$, and the refractive index in the medium $n_{im}$ in which the dyes are embedded (Eq. 3):

$$R_0 = \left(\frac{9ln(10)\kappa^2\Phi_{F,D}}{128\pi^5 N_A n_{im}^4} \int \tilde{F}_D(\lambda)\varepsilon_A(\lambda)\lambda^4 d\lambda\right)^{\frac{1}{6}} \qquad \text{(Eq. 3)}$$

where $N_A$ is Avogadro's number. The following section describes the factors that influence $R_0$ and FRET in detail:

<u>The extinction coefficient ($\varepsilon$):</u> One important parameter that feeds into the distance measurement is the extinction coefficient of the acceptor dye via its effects on the Förster distance and on the expected excitation rate in Alternating Laser Excitation / Pulsed Interleaved Excitation (ALEX/PIE) experiments. In the absence of an easy or affordable way to measure this parameter (it requires large amounts of dye for gravimetric analysis), the experimenter typically relies on the value given by the manufacturer. Fortunately, this factor does not usually vary depending on the environment of the fluorophores. However, as an example, the extinction coefficient of some of the popular organic dyes have been changed by the manufacturer in the past years a few times. Alternatively, the extinction coefficient of dyes may be theoretically assessed via the Strickler-Berg equation(Strickler and Berg, 1962), using the fluorescence quantum yield and lifetime. Since these quantities can be readily achieved from time-resolved fluorescence measurements, indirectly, or directly from nsALEX or PIE measurements, it might serve as an interesting route for attaining accurate values of the acceptor extinction coefficient.

<u>The fluorescence quantum yield ($\Phi_F$):</u> Another parameter that is often overlooked is the proper determination of $\Phi_F$. The $\Phi_F$ oftentimes changes upon labeling and can be sensitive to the labeling (local) position, and to the conformational state of the molecule. Even dyes that are considered relatively insensitive to their local environment have been shown to exhibit a large range in fluorescence QY upon conjugation to nucleic acids or proteins (e.g Cy3B with $\Phi_F$ of 0.8 or 0.4, respectively(Craggs *et al.*, 2019)) or even when labeling different nucleic acid bases with the same dye (e.g. Cy3B $\Phi_F$ values ranging from 0.19 to 0.97, over 12 different labeling positions on dsDNA, (Lerner, Ingargiola and Weiss, 2018)), leading to considerable variations in the values of $R_0$ (e.g. for the pair Cy3B-ATTO 647N, different labeling positions on dsDNA lead to values ranging between 54.8 Å and 64.5 Å(Lerner, Ingargiola and Weiss, 2018)). In addition, the fluorescence lifetimes and residual anisotropies of the two coupled dyes, Alexa488 and Alexa647, were analyzed for five distinct proteins with altogether 22 labeling sites(Peulen, Opanasyuk and Seidel, 2017). The fluorescence enhancement of cyanine-based Alexa647 correlated well with an increase of the dye's residual anisotropy that could potentially be indicative for dye species trapped on the protein surface. However, Alexa 647, a variant of Cy5, exhibits excited-state isomerization(Widengren *et al.*, 2001; White *et al.*, 2006). In such dyes, the rate of photo-isomerization influences their average fluorescence lifetimes, and this is, in many cases, correlated with a corresponding change in the fluorescence residual anisotropy(Sanborn *et al.*, 2007). Therefore, another possible interpretation for the observed differences in Alexa 647 fluorescence lifetimes correlated with the fluorescence residual anisotropies would be protein structural changes inducing change in the rate of photo-isomerization. In summary, independent determination of $\Phi_F$ for the different labelling positions(Lerner, Ingargiola and Weiss, 2018; Craggs *et al.*, 2019) or appropriate simulations are advisable.

Refractive index ($n_{im}$): Sometimes, an intermediate value of 1.4 is taken between the index of refraction of buffer (1.33) and that for proteins and DNA (~ 1.5). This approach reduces the maximal error in $R_0$ to ~4%(Clegg, 1992; Ingargiola, Segal, *et al.*, 2017). However, different values may be more appropriate depending on the geometry of the fluorophores. To date, the refractive index has received very little attention in the field(Knox and van Amerongen, 2002).

The dye orientation factor ($\kappa^2$): This parameter describes the relative orientation of the dyes and strongly depends on the model assumptions for dye mobility. Since in FRET the donor excitation energy is transferred to the acceptor after a few hundreds of picoseconds or a few nanoseconds, during which the dyes' orientations could change, the mean value of the orientational factor is typically taken. A well-known assumption that is often made is that the time it takes the rotation of the dyes about their bonds to reach isotropy is faster than the waiting time for the donor to transfer its excitation to the acceptor. For free rapidly rotating dyes, the orientation factor can be approximated by the isotropic mean value of $<\kappa^2>=2/3$. However, it may well be that one of the dyes is not freely rotating in timescales faster than the donor fluorescence lifetime (it may instead be interacting with its microenvironment). A method to estimate the lower and upper bounds for $<\kappa^2>$ from the donor and acceptor time-resolved anisotropies had been proposed already in the 1970's (Dale, Eisinger and Blumberg, 1979).  With this approach, it is possible to quantitatively estimate the value of $<\kappa^2>$, or at least to estimate the uncertainty in FRET due to uncertainty in $<\kappa^2>$. In smFRET measurements using the polarization resolved MFD modality, information on the donor and acceptor fluorescence intensities, lifetimes and anisotropies are collected simultaneously, and fluorescence anisotropy decays of different single-molecule populations can be used to assess the $<\kappa^2>$ uncertainty per conformational state. However, the range of possible $<\kappa^2>$ values could still be explained by a variety of different dye models. Several dye modeling approaches are now available to mitigate this potential pitfall(Muschielok *et al.*, 2008; Kalinin *et al.*, 2012; Beckers *et al.*, 2015; Dimura *et al.*, 2016). It is noteworthy that the majority of fluorophores used as donor and acceptor dyes in smFRET have exponential fluorescence decays, and hence have one major emission dipole. In these cases, the estimation of $<\kappa^2>$ depends on the orientation of these single dipoles. It has been proposed that the assumption of $<\kappa^2>=2/3$ would carry much less uncertainty, in cases where the fluorescence signal is emanating from more than one emission dipole, yielding non-exponential decays(Haas, Katchalski-Katzir and Steinberg, 1978). This is an intriguing idea that could help simplify the transformation of FRET efficiencies to distance information, taking into account a realistic estimation for $\kappa^2$.

Finally, we note that community-wide recommended routines to determine the Förster distance accurately are still under discussion.

## 3.2 Conversion from FRET efficiency into distances – The choice of a proper dye model

The Förster equation described above (Eq. 1) allows the extraction of a distance directly from a FRET efficiency measurement only if the positions and orientations of the donor and the acceptor molecules are constant. However, since dye molecules are typically attached to the macromolecules via flexible linkers, this is not the case even for a stable conformation of the macromolecule(Hellenkamp *et al.*, 2018; Ingargiola, Weiss and Lerner, 2018). Moreover, the Förster distance (Eq. 3) is not necessarily a constant, since slow rotational diffusion (nanoseconds or even slower) of the fluorophore about its linker may potentially lead to different $\kappa^2$ values in different energy transfer cycles(Eilert *et al.*, 2018).

To account for these problems, numerous different dye models have been proposed (Haas, Katchalski-Katzir and Steinberg, 1978; Muschielok *et al.*, 2008; Craggs and Kapanidis, 2012; Kalinin *et al.*, 2012; Beckers *et al.*, 2015; Dimura *et al.*, 2016; Schuler *et al.*, 2020) and tested experimentally(Woźniak *et al.*, 2008; Nagy, Eilert and Michaelis, 2017; Peulen, Opanasyuk and Seidel, 2017; Hellenkamp *et al.*, 2018). For any given FRET efficiency, different dye models will lead to different extracted distances; choosing an appropriate model (to the specific problem at hand) is therefore very important for accurate distance determination. However, since fluorescence emission as well as energy transfer can both be described by Poisson processes, even for fast rotational or translational diffusion (fluctuations in $R_{DA}$), there will be deviations in distances as well as relative orientations of the dye molecules on the photon by photon level. Averaging over these different situations is complicated due to the inherent non-linearity of the energy transfer process. In addition, since the linker lengths of typical dyes used in smFRET are not negligible (the equivalent of 5-6 carbons), rotational diffusion of dyes about their linkers lead to large changes in the inter-dye distance, $R_{DA}$. Such rotational dynamics sometimes occurs within times comparable to the fluorescence lifetime, which leads to changes in $R_{DA}$, from the moment of donor excitation, to the moment of the de-excitation, due to FRET. This is a well-documented phenomenon termed diffusion-enhanced FRET(Haas and Steinberg, 1984; Beechem and Haas, 1989; Orevi *et al.*, 2014), where the times in which photons were emitted report $R_{DA}$ biased to values shorter than in equilibrium (due to the increase in the probability for FRET to occur when $R_{DA}$ shortens)(Eilert *et al.*, 2018; Ingargiola, Weiss and Lerner, 2018). This has been treated by incorporating both rotational and translational diffusion as fluctuations of $R_{DA}$ (the inter-dye diffusion coefficient) inside a potential well in the reaction coordinate $R_{DA}$(Haas and Steinberg, 1984; Ingargiola, Weiss and Lerner, 2018). Similarly, when rotational motion changes the relative orientation of the donor and the acceptor, and therefore $\kappa^2$, this leads to changes in $R_0$. Therefore, energy transfer is also biased to favorable relative orientations of the donor and acceptor leading to a less well-known bias in $R_0$. To this end, a complete kinetic theory treating both rotational diffusion as well as translational diffusion has been developed(Eilert *et al.*, 2018). Interestingly, Monte-Carlo simulations show that the often applied simplifications such as the dynamic rotation - static translation model (i.e. $k_{rotation} \gg k_{FRET} \gg k_{diffusion} \gg k_{integration}$(Hellenkamp *et al.*, 2018)) can lead to distance errors, the magnitude of which depends on the donor fluorescence lifetime, the FRET efficiency as well as the two kinetic parameters namely, the dye molecules' diffusion constant and rotational correlation time.

Importantly, the uncertainty in $R_0$, the dye model and experimental precision is taken into account in integrative structural modeling by using carefully computed overall uncertainties in distance(Kalinin *et al.*, 2012; Dimura *et al.*, 2016; Peulen, Opanasyuk and Seidel, 2017; Hellenkamp *et al.*, 2018), so that these well-balanced FRET-restraints yielded structural models in benchmark studies that nicely agreed with models determined by X-ray crystallography.

### 3.3 The correction factor for detection efficiencies (the gamma factor)

The gamma factor in experiments on diffusing molecules: The most challenging correction in the analysis of intensity-based smFRET data accounts for imbalances in the donor (D) and acceptor (A) detection efficiencies and in the corresponding D and A quantum yields, the so-called γ-factor, and is necessary to compute an absolute FRET efficiency (Ha *et al.*, 1999). This factor corrects for the fact that the number of photons detected from the donor and acceptor fluorophores are not directly proportional to the number of their excitation/deexcitation cycles for two reasons: First, fluorophores, in general, have different fluorescence quantum yields. Second, the efficiency of collecting and detecting photons are different for the two channels due to different filters and optical transmission, the detector sensitivity to the different fluorophores and the fluorescence spectra of the dyes. Whenever a broad distribution of FRET efficiencies is available in the data set, the γ factors can be extracted from microsecond ALEX (μsALEX) or PIE data sets using the fact that the stoichiometry factor, S(Kapanidis *et al.*, 2004), is independent of FRET efficiency. However, it is essential that there are multiple species with different FRET efficiencies in the sample for such methods to work (Lee *et al.*, 2005). It is also necessary that the $\Phi_F$ values of the two dyes are identical for the different species. Whenever this is not available, accurate measurements of $\Phi_F$ of the dyes have to be performed for each species, and community-wide consistent routines need to be implemented to make these measurements simple. Fluorescence lifetime measurements and the correlated analysis of intensity and lifetime data could offer a solution to this problem(Rothwell *et al.*, 2003; Sisamakis *et al.*, 2010). When one or more species are dynamically averaged, a proper determination of the γ-factor becomes more challenging and different assumptions need to be made.

<u>The gamma factor in experiments on immobilized molecules:</u> When ALEX or PIE data are collected on immobilized samples, the γ-factor can be estimated using the stoichiometry and FRET efficiency information as discussed above, provided there is a significant distribution of FRET efficiency in the analyzed data set(Lee *et al.*, 2005). Additionally, the γ-factor can be estimated for individual molecules, where the acceptor fluorophore photobleaches first(Ha *et al.*, 1999; McCann *et al.*, 2010; Hildebrandt, Preus and Birkedal, 2015). Here, the decrease in the acceptor signal and the increase in donor signal can be directly compared. For this approach to be accurate, however, the acceptor must be photobleached rather than be in a transient (e.g. Redox) state that may potentially remain capable of absorption. The average γ-factor is then often applied to molecules where the donor photobleaches first. However, distributions of the γ-factor determined for individual molecules can be very broad, indicating some variability in its value.

### 3.4 Detection of dynamic averaging

Biomolecules are typically dynamic systems and conformational flexibility and dynamics, at short time scales are expected. A third challenge that needs to be addressed is the detection and analysis of these dynamically averaged populations (i.e. when the characteristic time of the conformational dynamics is shorter that the integration time needed to detect enough photons to calculate the FRET efficiency). A number of groups have developed methods for detecting and analyzing dynamic averaging. This includes analysis of the width of FRET efficiency distributions, investigation of the variance of single-molecule bursts, comparisons of average fluorescence lifetimes and intensity-based FRET efficiencies(Laurence *et al.*, 2005; Nir *et al.*, 2006; Dimura *et al.*, 2016; Ingargiola, Weiss and Lerner, 2018; Kilic *et al.*, 2018) or photon-based maximum likelihood methods(Pirchi *et al.*, 2016; Aviram *et al.*, 2018; Mazal *et al.*, 2019). When dynamic averaging is present, it becomes difficult to use the lifetime information to determine the γ-factor - however, fluorescence lifetime analysis can resolve conformations that exist on time-scales longer than the lifetime. It is also very difficult to check whether the quantum yield is identical in all conformations. The detection and analysis of dynamics is one of the issues addressed in the protein FRET challenge (in preparation) discussed below.

### 3.5 Improvements in dyes, detectors, sample handling and detection formats

Fluorescence dyes and their properties can strongly influence data quality in various fluorescence techniques. It was early recognized that dye photophysics play a pivotal role in smFRET experiments(Eggeling *et al.*, 1998; Kong *et al.*, 2007) and their correct interpretation. Optimization of dye photophysics can assist in avoiding photobleaching, saturation effects or photoblinking artifacts. The latter can introduce FRET-dynamics unrelated to conformational motion and, thus, potentially lead to misinterpretation of the data. Two strategies are used in dye development for optimal properties: (i) structural modifications of core dye structures (rhodamines, cyanines, oxazines, perylenes or others) aiming at higher absorption cross sections(Levitus, 2011), high fluorescence quantum yield(Grimm *et al.*, 2015), good chemical stability, water solubility and independence of the photophysical properties from the microenvironment(Levitus and Ranjit, 2011; Hell *et al.*, 2015; Michie *et al.*, 2017). (ii) Alternatively, photostabilizers can be used to reduce photodamage by triplet-states, oxygen and other reactive fluorophore species(Widengren *et al.*, 2007; Ha and Tinnefeld, 2012). Successful applications of photostabilizers, also in the context of smFRET, was achieved with Trolox(Rasnik, McKinney and Ha, 2006; Cordes, Vogelsang and Tinnefeld, 2009), β-mercaptoethanol(Campos *et al.*, 2011; Ha and Tinnefeld, 2012), ascorbic acid(Aitken, Marshall and Puglisi, 2008; Vogelsang *et al.*, 2008), cyclopolyenes(Targowski, Ziętek and Bączyński, 1987; Dave *et al.*, 2009), methylviologen(Vogelsang *et al.*, 2008), and a range of other compounds(Glembockyte, Lincoln and Cosa, 2015; Isselstein *et al.*, 2020). 'Self-healing' dyes, where the fluorophore is directly linked to a photostabilizing moiety, achieve high photon counting rates with intramolecular photostabilization(Liphardt, Liphardt and Lüttke, 1981; Altman *et al.*, 2012; van der Velde *et al.*, 2013; Juette *et al.*, 2014; Isselstein *et al.*, 2020). Utilization of switchable and/or multiple acceptors have also been recently suggested(Vogelsang, Cordes and Tinnefeld, 2009; Uphoff *et al.*, 2010; Krainer, Hartmann and Schlierf, 2015), caged and photoactivatable dyes(Jazi *et al.*, 2017), and such approaches could be further developed.

Arrays of single-photon avalanche diode detectors (SPAD arrays) and other novel detectors, coupled with novel optical detection geometries can increase smFRET measurement throughput and time resolution(Ingargiola, Segal, *et al.*, 2017; Gilboa *et al.*, 2019; Segal *et al.*, 2019). Microfluidics-based sample handling devices, including various mixers(Kim *et al.*, 2011; Wunderlich *et al.*, 2013), especially when coupled with detectors enabling multiplexing or multi-spot detection are expected to further improve measurement throughput, and allow automatic sample handling, as well as non-equilibrium measurements. All these additional dimensions point to the importance of precisely describing the components of an experimental setup used for any smFRET measurement, from optical elements (lenses, filters, etc.) to light sources and optomechanical/optoelectronical devices and their characteristics, detectors and their associated electronics, as they do contribute in many ways to the final recorded data, and cannot in general be inferred retrospectively.

## 4. Should steps towards a unified smFRET approach be taken?

*4.1 Argument in favor of a unified smFRET approach*

To demonstrate the reproducibility and reliability of smFRET measurements to the biological community, a multi-lab blind study of smFRET accuracy and precision has been performed. Twenty laboratories participated in measuring smFRET on several double-stranded DNA (dsDNA) constructs, exhibiting only low amplitude ultra-rapid dynamics, and hence considered relatively stiff (which is why they are often referred to as static)(Hellenkamp *et al.*, 2018). The quantitative assessment of the reproducibility of intensity-based smFRET measurements and of a unified data analysis and correction procedure(Lee *et al.*, 2005) was an important milestone, proving smFRET measurements to be consistent: studying six distinct samples with FRET pairs labeling different bases at different distances, the obtained mean FRET efficiencies by the 20 labs agreed within a range $\Delta E = \{0.02\text{-}0.05\}$. This effort established FRET-labeled dsDNA as consistent and reliable rulers for new optical setups as well as every day calibration, especially useful for new groups joining the community(Hellenkamp *et al.*, 2018).

Similarly, another multi-lab blind study of smFRET accuracy and precision for distance determination, using proteins undergoing ligand-induced conformational changes has been performed (in preparation). This study uses two distinct model proteins to assess the reproducibility and accuracy of smFRET measurements. In addition, for proteins stochastic labelling, storage, shipping and stability as well as dynamics and dye photophysics can complicate the analysis. The study also assesses the ability of smFRET to discover and quantify dynamics of transitions between different conformations at different timescales from seconds to microseconds.

Another challenge is the kinSoftChallenge2019 (http://www.kinsoftchallenge.com, *Schmid et al*, in preparation) that aims to evaluate existing tools for extracting kinetic information from single-molecule time trajectories (for time-binned or intensity-based measurements of immobilized molecules). This challenge aims to: (1) demonstrate the ability of smFRET-based kinetic analyses to accurately extract dynamic information from smFRET data; (2) provide the single-molecule (FRET) community a way to judge the different software tools out there; and (3) assess and communicate analysis procedures to yield reliable kinetic data and/or FRET state models. In the first part of this study, synthetic data was evaluated. In the second part of the study, experimental smFRET data are currently being evaluated.

One interesting outcome of the various multi-lab blind test challenges was the realization that the way data analysis and corrections are performed has a large impact on the final results. For example, the first smFRET consistency study(Hellenkamp *et al.*, 2018) led to recommendations on how to process raw data to obtain fluorescence intensities. Therefore, access to the raw data and the ability to process them with various analysis approaches is and will remain useful in moving the community forward. Currently, this is difficult as there are many variations in methods, their documentations, file formats and experimental procedures used in various laboratories. Furthermore, it is difficult to determine the optimal conditions, workflow and best practices even for existing, well-tested methods, since a comparison of these methods is time-consuming and the necessary information is, in many cases, not always available. Hence, open science practices as well as a platform for exchange of data and software would be beneficial.

One question that remains to be addressed is whether a user-unbiased, thus, fully automated analysis of smFRET data is possible. Therein, the FRET community needs to address the question how to overcome user bias in data analysis. Akin to structural models in the protein data bank (PDB) based on NMR and/or crystallization data(Rosato, Tejero and Montelione, 2013), smFRET benchmark criteria must be defined and eventually fulfilled to classify the results of FRET-based inter-dye distance measurements and kinetics (i.e. transition detection and state identification) as reliable. As an example, single-molecule video analysis usually suffers from a limited number of molecules (potentially bad statistics) and a potential user-bias, meanwhile the so-called molecular-sorting (i.e. picking the right molecules to be included in histogram and transition analysis). To overcome this, artificial intelligence, such as deep learning(Xu *et al.*, 2019) and machine learning(Stella *et al.*, 2018) approaches are being applied to overcome the user-bias in molecular sorting. In addition, an automated video processing and subsequent trace processing allows the fast identification of single-molecule trajectories and their characteristics(Preus *et al.*, 2015; Juette *et al.*, 2016; Börner *et al.*, 2018). A simplified and fully automated service for smFRET data analysis taking advantage of the community-driven challenges will not only reduce the user-bias tremendously and help to ease the analysis of smFRET data, but allow the application of smFRET to new fields and to further grow the community.

## 4.2 Arguments questioning the need for a unified smFRET approach

While this article covers an array of aspects related to establishing a smFRET community and recommendations on unifying smFRET practices for quantitative structural information, it is important to recognize that there is sometimes reticence regarding these ideas, and it is important to let these voices be heard.

The suggestions raised here, such as the appropriate use of methods, their use in the context of an open platform, benchmarking on simulations, providing the raw data, possibly in a standard file format that is easily read by others and proper documentation of preparative, measurements and analysis procedures, might merely be seen as what an exigent peer-reviewer would typically expect. This, in turn, raises the question of whether we need the effort and recommendations of the whole smFRET community for matters that should be obvious.

In addition, and in line with the multitude of existing software presented here (see Table 1), the question of who will determine their accuracy and utility in yielding similar datasets (for simulations) and results (for analyses) is also raised. Accompanying this ponderation are the questions: (i) how is the value of results that differ from some expected consensus taken into consideration, (ii) are 'expected results' chosen objectively? While section 4.1 (and the suggestions in this position paper) recommend certain guidelines, it is important to make sure that these stay as recommendations and will never evolve into 'requirements' disguised as good-practice guidelines.

The proposition of a modular software platform using algorithms that are accepted by the smFRET community, although it could be useful for many, might be unrealistic, mainly because (i) there are currently many analysis and software approaches, with as of yet little agreement on the algorithms, procedures and parameter values used, and hence such community-wide acceptance is premature; and (ii) there are many smFRET practitioners who are *de facto* members of the community, but do not necessarily wish to officially be part of the newly formed FRET community (http://fret.community).

Finally, it is important to remember that not all studies need the same degree of quantification and processing of smFRET data, depending on the aim of the study. Many smFRET studies use relative FRET efficiencies, rather than absolute values, to distinguish between conformational sub-populations and focus more on the transitions between them. Therefore, some of the requirements mentioned here might apply mostly to studies in which the sub-population-related FRET data is used in the modeling or reconstruction of the underlying biomolecular structure.

In summary, it is important to remember that as scientists, we should value independence of thought and creativity, and that any initiative that could be perceived as not encouraging such an attitude would be counterproductive. Still, the specific method used must be documented well so that the pathway from the sample, through measurement and analysis until interpretation can be followed and judged. Here we should indeed discuss the necessary information that should be provided together (for more details regarding smFRET for integrative structural biology, see section 6.6.).

## 5. Open Science practices

One of the cornerstones of the scientific method is the ability to reproduce experimental results. As experiments become more sophisticated, a clear description of experiments is crucial. Recent trends towards Open Science practices call for full transparency of the scientific process. Funding agencies embracing this philosophy (e.g. https://datascience.nih.gov/strategicplan) expect their grantees to publish in Open Access (OA) journals (and pay for the corresponding open access fees) or deposit manuscripts in repositories (e.g. Pubmed Central, arXiv, bioRxiv, ChemRxiv), deposit data (sometimes also raw data) in repositories (such as Zenodo, the Dryad Digital Repository, FigShare) as well as analyses codes (for instance in open notebook format in repositories such as GitHub), disseminate results and make them accessible for all. Open science disseminates knowledge by sharing results and the tools developed by independent scientists or teams working as part of a collaborative network. The smFRET community is committed to open science. Some tools are already in place, while other tools still need to be developed to make it easier to communicate smFRET accumulated and continuously growing knowledge and experience.

There is obviously some tension between the precepts of Open Science and requirements imposed by some intellectual property (IP) policies. IP rights, including patent laws, were put in place to promote the development of science and technology for the benefits of society, by allowing those developing intellectual property to retain rights regarding the use of the IP they develop. In fact, in some sense, patents were the first form of open access publication, only one with a restrictive license for reuse. We do not oppose intellectual property rights, but have misgivings about nondisclosure of methods, data and software. Other groups must be able to reproduce the analyses of existing data (and extend upon them) and, if needed, be able to reproduce experiments having produced these data, for confirmation purposes. The acquisition and analysis must be modifiable and extendable as the end users see fit, in agreement with the license chosen by the data or software creator. This license should be set as liberal as possible, taken into account IP considerations mentioned above, but also encourage recognition of the sometime enormous effort invested in producing successful protocols, designs, data or software. Ultimately, if practiced fairly, open science should entice everyone, including commercial vendors to adopt and contribute to community-defined file formats, provide free file conversion codes and open their analysis tools for scrutiny by the community.

Finally, online and public repositories are a form of data and knowledge backup, which most of us have learned the hard way, is difficult to achieve and maintain at the scale of a single laboratory.

## 6. 'Soft recommendations' for smFRET measurement and analysis practices for quantitative structural studies

For smFRET to reach its full potential for quantitative structural studies, many of us believe that it would be beneficial for the community of practitioners to develop general and flexible set of 'soft recommendations' for data collection, analysis and sharing practices.

### 6.1 What are the current difficulties and challenges?

Beyond the challenges addressed in section 3, we want to point out additional technical challenges. In smFRET of freely-diffusing single molecules, the raw data includes a sequence of photon detection times from (at least) two detectors that register both photons from the fluorescent analyte and background photons. The first step includes separating the fluorescence photons from the background and identifying single-molecule photon bursts. Afterwards, bursts are selected according to thresholds, based on different features of the photon bursts (e.g. size, duration, brightness). Since the threshold values are in many cases chosen arbitrarily, it might have an impact on the resulting smFRET histogram. In addition, the experimental setup (e.g. filters and dichroics used, excitation and detection characteristics) might, in some cases, also affect the results. In many cases, different measurements require using different burst search and filtration parameter values. Hence, already the first step in the analysis is context-dependent. Similarly, for TIRF-based measurements, there are various protocols for extracting the fluorescence intensity of the donor and acceptor signals and the corresponding background(Preus, Hildebrandt and Birkedal, 2016).

For both confocal- and TIRF-based smFRET measurements, the second step is to correct the data for donor fluorescence bleed-through to the acceptor fluorescence detection channel, acceptor direct excitation (rather than via FRET), the γ factor describing the differences in donor and acceptor fluorescence quantum yields and detection efficiencies and other correction factors. There are currently different ways to characterize these correction factors (see discussion in section 3.3). This is yet another example of the context-dependence of different procedures in the analysis of smFRET data. We recommend that a rigorous study is performed to compare intensity- and lifetime-based smFRET measurements on well-characterized model systems, to verify which information is best obtained from intensity or lifetime information, assess the consistency of the different methods to determine the needed correction factors and the identify potential pitfalls in the correction procedure.

The final goal of these steps is to identify sub-populations of bursts with common FRET efficiency values, better known as FRET sub-populations. While ideally these sub-populations would represent the mean FRET efficiency values of conformational states, often these are merely a result of time-averaging over several different conformational states, interconverting faster than the typical durations of single-molecule photon bursts (a few milliseconds). There are currently various methods for analyzing photon statistics to identify and quantify the exchange between conformational states within bursts(Felekyan *et al.*, 2013; Lerner *et al.*, 2018) the choice between them and the way they are used may sometimes be conceived as arbitrary (see detailed discussion about dynamics in section 2.2).

The multi-laboratory efforts for quantitative smFRET analyses, both for DNA(Hellenkamp *et al.*, 2018) and for proteins (*in preparation*) have been very productive, and advocative of openness, and we encourage researchers in the field to post the raw data and analysis codes on public repositories. This notwithstanding, there are three steps to transform FRET measurements to structures: (i) transforming the FRET data, in general *D*, of conformational sub-population to inter-dye distances, (ii) transforming the structures, referred to generally as *M*, to inter-dye distances by an appropriate dye model(Beckers *et al.*, 2015; Dimura *et al.*, 2016; Steffen, Sigel and Börner, 2016) , and (iii) computing the data likelihood of structures and the given data, $L(M|D)$. Different laboratories use different approaches and all steps necessary for the transition from FRET information to distance information are still under debate. Consequently, different laboratories employ different approaches to analyze their experimental results in context-dependent manners.

To bridge the different approaches between members of the FRET community, we suggest a concerted action for an appropriate use of methods in a context of an open platform with many use cases. These include providing results of simulations that explain the choices of certain stages in the analysis procedure, providing the raw data in a standard file format easily readable by all FRET practitioners and properly documenting the whole analysis procedure together with its code. Additionally, reporting best practices does not include only how data was analyzed, but also how materials were prepared, purified and characterized, and whether dye-labeled biomolecules under measurement are functioning in a manner that reflects that of the unlabeled, wild-type biomolecule(Orevi *et al.*, 2014; Best *et al.*, 2018; Lerner, Ingargiola and Weiss, 2018; Riback *et al.*, 2019). By accumulating the experience and expertise of the community, we aim to provide recommendations and guidelines that help attain reliable and reproducible results while keeping the final approach choices open to the individual scientists.

## 6.2 Consistent reporting of preparative practices

For studying biomolecular conformations and their dynamics by smFRET, biomolecules of interest must be labeled with dyes that are suitable for single-molecule fluorescence detection because the intrinsic fluorophores are not stable enough and absorb in the UV. These dyes usually include three units: (i) a chemically reactive group that forms a covalent bond preferentially with a specific type of moiety/residue (in a DNA base or in an amino acid, respectively), (ii) a linker of a few connecting bonds and (iii) a pi-conjugated fluorophore that typically has hydrophobic regions, could be fully or partially planar and is often bulky. When measuring an intra-molecular distance within a biomolecule, smFRET requires conjugating two dye molecules. Additionally, site-specific conjugations in proteins require introducing point mutations that will accommodate the specific conjugation chemistry to be used. It is obvious that conjugation of a dye molecule to a protein or DNA and the features mentioned above may introduce both structural and functional perturbations, relative to the unlabeled biomolecule(Enderlein *et al.*, 2005; Borgia *et al.*, 2016). Therefore, it is important not only to report on the labeling procedures, the purification procedures of its products and the labeling efficiency, but also (biochemical) control experiments to determine the extent to which the dye-labeled biomolecules represent the wild-type behavior by means of structure (e.g. secondary structure content using far-UV circular dichroism, CD, dynamic light scattering, DLS, small-angle X-ray scattering, SAXS), thermodynamic stability (e.g. thermally-induced transition curves) and biological activity(Orevi *et al.*, 2014; Borgia *et al.*, 2016; Lerner, Ingargiola and Weiss, 2018).

Single-molecule experiments have the advantage that each molecule is observed individually and various criteria can be used to decide which molecules are included for further analysis. Although single-molecule experiments are often advertised as being able to detect rare events, a sufficient amount of statistics and additional control experiments (e.g. comparison with mutations, or with partially or non-functional ligands or substrates) are required to ensure that the detected sub-population is biologically-relevant and not an anomaly. Hence, a robust smFRET analysis should use optimized labelling protocols resulting in products labeled with both donor and acceptor dyes at substantial fractions (a minimum of 10%, but higher labeling fractions are preferred), and rigorous quality control of purity (e.g. analytical chromatography, mass spectrometry). Only optimally-labeled samples enable statistically-relevant and meaningful information to be collected, and prevents "cherry-picking" of individual molecules. In principle, proper analytical chromatography is performed to achieve >90% pure labeled samples that are separated from the free dyes and from the un-labeled samples, are recommended. A possible recommended smFRET protocol is to start with analysis of diffusing fluorescent species to determine the (i) quality of labelling, (ii) number and (iii) FRET properties of major biochemical species. With this information at hand, smFRET analysis of surface-immobilized molecules can be performed, where the functionality of biomolecules dual-labeled with donor and acceptor dyes was verified by comparison of the percentage of dual-labeled biomolecules with biochemical activity of the labeled species beforehand (taking into account the possible fraction of un-labeled species in solution). The potential of user bias in selection of fluorescent time traces might be less probable, if guided by the information obtained from the robust diffusion-based analysis. This two-step process also allows the assessment of whether biomolecule-surface interactions interfere with the biochemical activity, which would reveal different results obtained in step one (diffusion-based smFRET) and step two (smFRET with immobilized molecules).

Sample immobilization via the surface attachment of biomolecules must be carefully performed in order to systematically eliminate spurious contributions from molecules that are non-specifically bound(Lamichhane *et al.*, 2010; Traeger and Schwartz, 2017). Extensive efforts have been made in the field to optimize surface passivation procedures to address this potential issue. Alternatively, molecules can be encapsulated in liposomes(Boukobza, Sonnenfeld and Haran, 2001; Okumus *et al.*, 2004). However, even here, the extrusion process and the fact that not all proteins end up inside the liposome can also reduce the fraction of functioning proteins. In addition, interactions between the protein and/or dyes and the lipids can pose a problem. Control experiments demonstrating the specific nature of the surface immobilization strategy are therefore paramount. Despite these difficulties, much can be learned from surface-based experiments. Nevertheless, it is important that the conditions and percentage of functional or dynamic molecules be openly described within the publication(Roy, Hohng and Ha, 2008; Lamichhane *et al.*, 2010). Other elegant integrated approaches such as DNA-origami platforms(Gietl *et al.*, 2012; Bartnik *et al.*, 2020) or the use of anti-brownian electrokinetic (ABEL) traps(Cohen and Moerner, 2005), or nanochannel devices(Tyagi *et al.*, 2014; Fontana *et al.*, 2019) can also be used to exclude such effects.

Besides testing biochemical parameters and checking for correct protein function, additional artifacts of the dyes need to be taken into account. These can be photoblinking and photobleaching, which both create artifactual FRET-species when not properly recognized(van der Velde *et al.*, 2016) or saturation effects (reducing overall observed brightness)(Nettels *et al.*, 2015) occurring when e.g., acceptors are used that have strong tendency for triplet-state formation or photoisomerization, or artificial high-FRET states due to dye-dye interactions(Sánchez-Rico *et al.*, 2017). Hence, we recommend validating all important biological results with a second FRET pair (and perhaps more)(Voelz *et al.*, 2012; Borgia *et al.*, 2016, 2018; Lerner *et al.*, 2017; Husada *et al.*, 2018; de Boer *et al.*, 2019). Molecules are also often labeled stochastically using double cysteine mutations, in which case a mixture of donor-acceptor and acceptor-donor labeled molecules is measured. This might cause problems when the donor/acceptor dyes experience different microenvironments at the different dye labeling positions, leading to different $\Phi_F$ values. This potential problem is avoided by applying orthogonal labeling approaches using unnatural amino acids(Milles, Tyagi, *et al.*, 2012; Sadoine *et al.*, 2017; Quast *et al.*, 2019). This can also be avoided by proper planning of double cysteine labeling, either via differences in thiolate reactivities(Jacob *et al.*, 2005; Orevi *et al.*, 2014) or via different chromatographic elution profiles of different dye-labeling species(Orevi *et al.*, 2014; Zosel, Holla and Schuler, 2020). It is important to note that spectroscopic parameters can also be used for validation (i) through the analysis of the fluorescence anisotropies of donor and acceptor fluorophores(Rothwell *et al.*, 2003; Hellenkamp *et al.*, 2017) and (ii) using the consistency of the FRET data in the network through the analysis of position-specific deviation (weighted residuals) of inter-dye distance between the structure model and the experiment(Dimura *et al.*, 2019). When the local environment influences the photophysical properties of either the donor or acceptor fluorophore, different FRET subpopulations will be erroneously detected. However, through properly designed control experiments (for example, by preparing single cysteine mutants for both positions and both dyes) such errors could, in principle, be corrected.

As a general note, a rigorous screening procedure should be developed for checking whether the mutations made to a biomolecule for labeling and/or the labels themselves influence the functionality of the biomolecule. In principle, dye-labeled biomolecules that exhibit biochemical differences (catalytic activity, binding affinity, thermodynamic stability etc.) beyond a certain threshold value as compared to the wild-type (*wt*) counterpart, should be discarded and not used. However, what tests should be performed and what threshold values used? Could there be guidelines or recommendations that will help diminish arbitrary or subjective choices? In addition, much can still be learned about the functionality of a protein and the molecular mechanisms behind its function, even when the dyes alter the system (when no better labeling alternatives are available). Therefore, we believe all of the above characterizations should be documented and provided not only in the preparative section of a paper, but also when reporting the results. In the future, when smFRET results will be part of wwPDB accession codes, these reports should also be given.

### 6.3 The need to simulate measurements

When developing new analysis tools, it is necessary to know the capabilities and limitations of the developed method. In the end, one would like to know to what extent the data analysis procedure used by a given team, represents the measured system. This question is relevant, especially when other common analysis procedures yield different results(Blanco and Walter, 2010; Beckers *et al.*, 2015; Chen *et al.*, 2016). Using prior knowledge about the measured system and about the experimental setup, one can perform simulations of the measurements and then check whether the different analysis procedures again retrieve different results, when the ground-truth is the same. This way, it is possible to provide proper reasoning for one choice of an analysis procedure compared to others, at least in a given context.

As an example, for confocal-based FRET of freely-diffusing single molecules, many simulation software packages have been developed. Common to all approaches is the simulation of the Brownian motion of particles in a box and the evaluation of the fluorescence emission based on the excitation-emission profile, generally assumed to be Gaussian. This results in a trajectory of detected photons from each molecule as a function of time, and hence, as a function of its position at each moment. All photons from all times and all molecules together with an additional background process, form a dataset that simulates confocal-based smFRET experimental raw data. Then, the analysis of the simulated smFRET data can be performed using different procedures and tested against the known ground-truth. When such an approach is adopted, different teams could keep using their own analysis procedures yielding results that are comparable across different laboratories. Currently, there are several simulation packages available: (i) PyBroMo (ii) module within the PAM software package, (iii) Burbulator, (iv) simFCS (https://www.lfd.uci.edu/globals/) and (v) a module in Fretica (https://www.bioc.uzh.ch/schuler/programs.html). For the simulation of immobilized single-molecules and single-molecule videos (SMV), there is a module in the software package MASH-FRET available(Börner *et al.*, 2018).

PyBroMo (https://github.com/tritemio/PyBroMo) is a Python-based software package for simulating freely-diffusing single-molecule fluorescence detection including Brownian motion and dynamical exchange currently between two-states. This approach has been employed in a few recent works(Ingargiola, Lerner, *et al.*, 2017; Lerner, Ingargiola and Weiss, 2018; Hagai and Lerner, 2019). Analogous approaches, including the full simulation of fluorescence lifetime and anisotropy information as well as conformational multi-state (up to 8 states) exchange dynamics, are implemented in the MATLAB-based PAM software package ([https://www.cup.uni-muenchen.de/pc/lamb/software/pam.html](https://www.cup.uni-muenchen.de/pc/lamb/software/pam.html)) and the LabView-based Burbulator software(Kalinin *et al.*, 2010; Felekyan *et al.*, 2012; Dimura *et al.*, 2016; Barth, Voith von Voithenberg and Lamb, 2019) ([http://www.mpc.hhu.de/software/software-package.html](http://www.mpc.hhu.de/software/software-package.html)), and have been applied to benchmark novel quantitative analysis methods to obtain structural and kinetic information. Fret-ica ([https://www.bioc.uzh.ch/schuler/programs.html](https://www.bioc.uzh.ch/schuler/programs.html)) enables the simulation of single-molecule multichannel-detection of immobilized molecules and mixtures of freely diffusing species, including dynamic exchange between an arbitrary number of internal (e.g. conformational) states. The simulation of fluorescence lifetime and anisotropy information can be included(König *et al.*, 2015; Zosel *et al.*, 2018). The Matlab-based MASH-FRET software package ([https://rna-fretools.github.io/MASH-FRET/](https://rna-fretools.github.io/MASH-FRET/)) has been applied to evaluate transition detection and state identification algorithms used in particular for time-binned smFRET trajectories(Hadzic *et al.*, 2018) as well as for spot detection in single-molecule videos (SMV).

To ensure simulations can be shared without the need for exchanging large files of simulated photon counts or single-molecule videos, the PyBroMo, PAM, Burbulator and MASH-FRET packages provide initialization files that contain the simulation parameters and the seed of the random number generator, so that identical trajectories can be produced(Dimura *et al.*, 2016). In the absence of simulated data, having data measured and analyzed by various groups openly available is also a possibility to check various analysis approaches.

### 6.4 Standard file format

To expedite the exchange of data between different groups and testing of different analysis methods, it would be beneficial to have as few standard file formats as possible, to avoid the multiplication of ad hoc formats developed independently and requiring as many software codes to be effectively shared with the community. In fact, the most vexing issue with the absence of recommendations for 'good practice' is that individual labs themselves go through cycles of booms and busts as far as data and file formats are concerned, where one student developing a format (and accompanying software) that is used for a while, before it is superseded a few years later (or in parallel) by a different combination when a different student takes over. A vast amount of data can thus become rapidly obsolete and inaccessible because of lack of documentation and or support. Online data deposition in well-documented file formats would therefore save a lot of headache to many laboratories.

The reuse of smFRET data could also be performed online as part of the deposition of data in databases. To do so, the raw experimental data should be supplied in a universal data file format that could be easily read and scrutinized. Ideally, the file should store both raw data and sufficient metadata such that it will completely specify the measurement, setup and sample. Such a package (data and metadata) allows third parties to reanalyze the data in order to confirm published results or to benchmark new methods. Ideally, the metadata should be stored in a human-readable text-based format, while space-efficient storage of the raw photon data should be ensured by lossless compression. To promote the adaptation of new file formats, conversion tools for older file formats should be provided so that future software codes can focus on handling one (or at least only a very few) common file format.

There are nowadays many different file formats, developed by different research groups and companies that perform similar types of experiments. Some of these file formats have been defined (and luckily often made publicly available) by companies that develop components for experimental setups, or the whole instrument. While they are supported in their corresponding commercial software, they are, in general, closed and therefore not fulfilling the goal of open science. Moreover, they are not guaranteed to be perennial, which poses an additional challenge to the community.

<u>File format for (time-correlated) photon counting with point-detector data:</u> originally, a multiscale counting unit with an external clock was used to record the number of detected photons (photoelectrons) in a defined time interval to compute count rates for the fluorescence intensity(Rigler *et al.*, 1993). To increase the time resolution, the time to the preceding signal photon was recorded as a measure of the macroscopic detection time of the events in the experiment (macrotime) and defines the basic single-photon-counting format. To add fluorescence lifetimes to the previous measurement, pulsed excitation together with time-correlated single photon counting (TCSPC) was combined to record the photon detection time relative to the excitation laser pulse moment (microtime) with no loss of time information(Brooks Shera *et al.*, 1990; Tellinghuisen *et al.*, 1994; Brand *et al.*, 1997; Schaffer *et al.*, 1999; Eggeling *et al.*, 2001). The accumulation of microtime data encodes the information about the fluorescence lifetime. In 2005, the company Becker & Hickl (B&H) introduced synchronized and deadtime-free TCSPC electronics(Felekyan *et al.*, 2005) so that the microtime and macrotime can be linked by using the excitation laser as an internal clock. This way, the experiment time can be recorded for up to hours with picosecond resolution and a full fluorescence correlation from picoseconds to thousands of seconds can be calculated(Felekyan *et al.*, 2005). As more than one detector is used in FRET and in polarization-resolved studies(Widengren *et al.*, 2006; Wahl *et al.*, 2008), each time stamp for a detected photon is stored together with the information for the specific detector that registered it (Figure 3). This format has been widely-adopted by commercial companies (e.g. B&H, PicoQuant, PQ) for their TCSPC electronics, which is often used for solution and imaging studies with point detectors.

These basic principles have recently been extended in the Photon-HDF5(Antonino Ingargiola *et al.*, 2016) file format that connects the required metadata with the raw photon information in a single, space-efficient format that is suitable for sharing and long-term data archival. From the beginning, the development of the format has been done openly on GitHub (https://github.com/Photon-HDF5), encouraging comments and involvement from the community. The rich metadata included in Photon-HDF5 promotes best practices in documenting experimental details. To better communicate data transfer using the Photon-HDF5 file format, scripts for conversion of multiple different file formats (e.g. from Becker & Hickl and PicoQuant) into the Photon-HDF5 file format have been introduced. Photon-HDF5 might not be the ultimate format and might need further development, but its underlying philosophy is well-aligned with the proposed recommendation in this article.

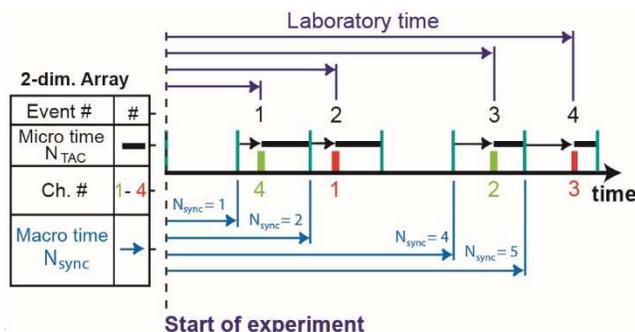

**Fig. 3:** The format of the time-tagged time-resolved description of single-photon detections, used in smFRET. Each photon that is detected by one of ≥1 detectors (Ch. #), is detected at a given time (laboratory time), comprised of the addition of some multiple of the detection clock time (yielding the photon macrotime) and the time relative to the moment of excitation (yielding the photon nanotime). The representation of the photon detection time in steady-state or time-resolved smFRET is by the macrotime in time-tagged (TT) format or by the sum of the macrotime and the nanotime in time-tagged time-resolved (TTTR) format, respectively. (Fig. 3 is adaped from (Felekyan *et al.*, 2005), with permission).

<u>File format for wide-field/TIRF camera-based acquisition:</u> This type of data is acquired as a stack of images (e.g. TIFF). To extract time trajectories quite a bit of selection/filtering is needed in order to yield timed-binned photon lists (spot identification; D and A spot association; thresholding; time trajectory extraction; background subtraction, etc.). A file format for TIRF-based smFRET (immobilized) measurements has been pro-

posed(Greenfeld *et al.*, 2015). Alternatively, human-readable plain text files - with an agreed format - work well for TIRF-based smFRET traces, since the amount of data is small, in particular when compared to the raw TIFF stacks. Such plain text files have been used in the kinSoftChallenge successfully, by several labs.

Exchange file format for processed data: Currently, multiple software packages exist for the analysis of (sm)FRET experiments, each with its strengths and weaknesses. While the software packages are often modular, the use of a certain analysis tool often requires that the data be reanalyzed in the given software. To address this situation, we recommend that exchange file formats be defined for different levels of processed data (e.g. burst analysis datasets, FCS curves, FRET efficiency histograms, TCSPC decays, photon counts for photon distribution analysis). In this way, researchers will be able to combine different software packages in a modular workflow tailored to their specific use case. Exchange file formats will also lower the barrier to adopt new analysis approaches, ensuring that the full capabilities of the tools available in the community can be implemented with minimal time investment and that the maximum amount of information (per the type of processed data) can be extracted from the experimental data. In addition, if processed data is stored with publications, the use of agreed-upon file formats will enable researchers to reproduce the analysis with minimal effort. The deposition of processed data with publications will also be useful for the development of forward-modeling approaches that aim to reproduce the experimental information (fluorescence decays, FRET efficiency histograms) from the structural models or ensembles(Köfinger *et al.*, 2019). To facilitate the exchange between different software packages and the adoption of exchange file formats, we propose that a command-line file conversion tool should be developed (https://github.com/Fluorescence-Tools/exchange-formats). However, we believe it is important to keep some unique identifier in each format exchange operation in order not to compromise the concept of raw data. To fulfill this requirement, it is crucial to properly document metadata content referring to the raw data from which the processed data was extracted.

## 6.5 The need for proper documentation of data analysis practices

To realize the FAIR principle that data should be "Findable, Accessible, Interoperable, and Re-usable", the procedures taken to analyze experimental data should not only be provided, but also well-documented. That includes describing each step, the reasons for taking it and its meanings. To ensure that the analysis remains transparent and tractable, all parameters and settings used for analysis should be stored at any point, i.e. analysis input and output files should accompany the data.

There is already a large number of different freely available programs (see Table 1) that offer a large variety of analysis procedures for single-molecule photon trajectories (confocal) and single-molecule videos (widefield/TIRF) data. Depending on the user community and user experience, graphical and command line workspaces were realized. In this context, we strongly believe and recommend that the analysis codes should be open for the community to read as well as write and modify. To improve the inter-operation between methods and to establish convenient documentation protocols, it is essential to work in an open multivalent environment. Establishing a community-wide working group "*Analysis software for FRET*" in the context of the FRET community might assist in organizing and moderating this process. However, the product of this working group should be treated as 'soft recommendations' and should not, at any moment, impose new software developments or to use any of the already existing tools.

For this goal, the use of the browser-based software such as "*Jupyter* notebooks", and/or other available workspaces, may serve as a potentially convenient platform for experts developing methods and depositing documented analysis procedures. Such workspaces provide an interactive scripting environment by combining a rich document with code commands as well as code outputs (e.g. figures, tables, comments, equations) and explanations in a single web-based document. Such web-based workspace environments can be easily read and distributed over the internet as well as re-run, checked or modified and supports several programming languages including python, R, C++ and, to some extent, MATLAB. In the development stage, we suggest creating a separate library to implement all the core functions and steps for a given method. These functions can be called by a specific analysis workflow allowing the user to change parameters and explore results. With this approach, the library can be developed using modern software engineering approaches and tooling, thus minimizing the exchange of bugs and facilitating maintenance. Software engineering approaches to be used in scientific software include version control, code review, unit testing, continuous integration and auto-generation of HTML manuals. In the next step, well-documented, easy-to-use *Jupyter* notebooks, can help newcomers to the field perform such complex analyses already in the notebook environment, with minimal adaptation efforts. Indeed, well-documented, easy-to-use notebooks, including such complex analysis tasks, have been provided to and by the community (for example for assessing smFRET dynamics in microsecond alternating laser excitation, μsALEX(Kapanidis *et al.*, 2005; Lee *et al.*, 2005), measurements(Lerner, 2020) or in nanosecond alternating laser excitation, nsALEX(Laurence *et al.*, 2005), pulsed-interleaved excitation, PIE(Müller *et al.*, 2005), measurements(Lerner, 2019)). Other software have recently started following this route as well, providing a suite of notebooks (based on FRETBursts(A Ingargiola *et al.*, 2016) and available on the smfBox(Ambrose *et al.*, 2019) GitHub, https://craggslab.github.io/smfBox/) for the workflow from raw photon-HDF5 files(Antonino Ingargiola *et al.*, 2016) through the determination of all correction factors, arriving at absolute, accurate FRET efficiencies.

Although the notebook approach offers advantages to software developers and experienced users, it might be difficult for many end-users to navigate the command-line environment and adapt to the script-based workflow. For such users, it is safer and more convenient to use established and tested algorithms embedded in graphical user interfaces (GUIs). Indeed, there is a large variety of user-friendly software available (Table 1). As a first step towards this goal, we propose that a software library (e.g. a Python package) should be established that contains established and tested algorithms for the analysis of fluorescence experiments, allowing them to be efficiently distributed and implemented in existing notebooks. Such efforts have already been initiated in the *FRETbursts software package*(A Ingargiola *et al.*, 2016) and a GitHub group has been established at https://github.com/Fluorescence-Tools to collect software packages and connect software developers.

**Table 1: List of software packages for the analysis of FRET experiments and integrative FRET-restrained structure modeling**

| Software | Type | Description | URL |
|---|---|---|---|
| OpenSMFS | Confocal | A collection of tools(A Ingargiola *et al.*, 2016) for solution based single molecule fluorescence spectroscopy, including single-molecule FRET, FCS, MC-DEPI(Ingargiola, Weiss and Lerner, 2018). | https://github.com/OpenSMFS |
| MFD Spectroscopy and Imaging | Confocal | A software package for confocal fluorescence spectroscopy and imaging experiments using multiparameter fluorescence detection (MFD) with all tools (fFCS, PDA, seTCSPC, Burbulator) and multiparameter fluorescence image spectroscopy (MFIS).(Kühnemuth and Seidel, 2001; Felekyan *et al.*, 2005; Antonik *et al.*, 2006; Widengren *et al.*, 2006) | http://www.mpc.hhu.de/software/software-package.html |
| H²MM | Confocal | H²MM is a maximum likelihood estimation algorithm for photon-by-photon analysis of single-molecule FRET experiments(Pirchi *et al.*, 2016). | http://pubs.acs.org/doi/suppl/10.1021/acs.jpcb.6b10726/suppl_file/jp6b10726_si_002.zip |
| ALiX | Confocal | ALiX is developed for basic research on diffusing two-color single-molecule FRET in single or multiple spot geometries.(Ingargiola, Lerner, *et al.*, 2017) | https://sites.google.com/a/g.ucla.edu/alix/ |

| smfBox | Confocal | Confocal smFRET platform, providing build instructions and open-source acquisition software.(Ambrose *et al.*, 2019) | https://craggslab.github.io/smf-Box/ |
|---|---|---|---|
| ChiSurf | Confocal Ensemble Modelling | ChiSurf is a fluorescence analysis platform for the analysis of time-resolved fluorescence decays.(Peulen, Opanasyuk and Seidel, 2017) | https://github.com/Fluores-cence-Tools/ChiSurf/wiki |
| rFRET | Confocal Imaging | rFRET is a comprehensive, Matlab-based program for analyzing ratiometric microscopic FRET experiments.(Nagy *et al.*, 2016) | https://peter-nagy.webs.com/fret#rfret |
| PAM - PIE Analysis with MATLAB | Confocal Imaging Ensemble | PAM (PIE analysis with Matlab) is a GUI-based software package for the analysis of fluorescence experiments and supports a large number of analysis methods ranging from single-molecule methods to imaging.(Schrimpf *et al.*, 2018) | https://www.cup.uni-muenchen.de/pc/lamb/soft-ware/pam.html |
| Fretica | Confocal | Fretica, a Mathematica package with a backend written in C++, is a user-extendable toolbox that supports MFD, PIE/ALEX, PCH(Müller, Chen and Gratton, 2000; Huang, Perroud and Zare, 2004), FIDA(Kask *et al.*, 1999; Gopich and Szabo, 2005), PDA(Antonik *et al.*, 2006; Ernst *et al.*, 2020), recurrence analysis(Hoffmann *et al.*, 2011), fluorescence lifetime fitting, FLIM, FCS, FLCS(Dertinger *et al.*, 2007; Arbour and Enderlein, 2010), dual-focus FCS, nsFCS(Net-tels *et al.*, 2007; Schuler and Hofmann, 2013), maximum likelihood estimation from photon-by-photon(Gopich and Szabo, 2009) and binned trajectories, simulation of confocal experiments and more. | https://www.bioc.uzh.ch/schuler/programs.html |
| MASH-FRET | TIRF | MASH-FRET is a Matlab-based software package for the simulation(Börner *et al.*, 2018) and analysis of single-molecule FRET videos and trajectories (video processing(Hadzic *et al.*, 2016), histogram analysis(König *et al.*, 2013) and transitions analysis(König *et al.*, 2013; Hadzic *et al.*, 2018)). | https://rna-fretools.github.io/MASH-FRET/ |
| miCUBE | TIRF | TIRF smFRET platform, providing detailed build instructions and open-source acquisition software. (Martens *et al.*, 2019) | https://hohlbein-lab.github.io/miCube/index.html |
| TwoTone | TIRF | A TIRF-FRET analysis package for the automatic analysis of single-molecule FRET movies.(Holden, Uphoff and Kapanidis, 2011) | https://groups.phys-ics.ox.ac.uk/genema-chines/group/Main.Soft-ware.html |
| ebFRET | TIRF | ebFRET performs combined analysis on multiple single-molecule FRET time series to learn a set of rates and states.(van de Meent *et al.*, 2014) | https://ebfret.github.io/ |
| vbFRET | TIRF | vbFRET uses variational Bayesian inference to learn hidden Markov models from individual, single-molecule fluorescence resonance energy transfer efficiency time trajectories.(Bronson *et al.*, 2009) | http://www.colum-bia.edu/cu/chemis-try/groups/gonzalez/soft-ware.html |
| STaSI | TIRF | STaSI uses the Student's t test and groups the segments into states by hierarchical clustering. (Shuang *et al.*, 2014) | https://github.com/Lande-sLab/STaSI |
| iSMS | TIRF | iSMS is a user-interfaced software package for smFRET data analysis. . It includes extraction of time-traces from movies, traces grouping/selection tools according to defined criteria, application of corrections, data visualization and analysis with hidden Markov modeling and import/export possibilities in different formats for data sharing.(Preus *et al.*, 2015) | http://isms.au.dk/ |
| HaMMy | TIRF | smFRET analysis and hidden Markov modeling.(McKinney, Joo and Ha, 2006) | http://ha.med.jhmi.edu/re-sources/ |

| smCamera | TIRF | smFRET data acquisition (Windows .exe) and analysis (IDL, MATLAB) with example data.(Roy, Hohng and Ha, 2008) | http://ha.med.jhmi.edu/re-sources/ |
|---|---|---|---|
| SMACKS | TIRF | SMACKS (single molecule analysis of complex kinetic sequences) is a maximum-likelihood approach to extract kinetic rate models from noisy single molecule data.(Schmid, Götz and Hugel, 2016) | https://www.singlemole-cule.uni-freiburg.de/soft-ware/smacks |
| SPARTAN | TIRF | Automated analysis of smFRET multiple single molecule recordings. Includes extraction of traces from movies, selection of traces accord-ing to defined criteria, application of corrections, hidden Markov modeling, simulations, and data visualization. (Juette et al., 2016) | https://www.scottcblanchard-lab.com/software |
| FPS | Modeling | A toolkit for Förster resonance energy transfer (FRET) restrained modeling of biomolecules and their complexes for quantitative appli-cations in structural biology.(Kalinin et al., 2012) | http://www.mpc.hhu.de/soft-ware/fps.html |
| Fast NPS | Modeling | A nano-positioning system for macromolecular structural analysis (to be extended by the authors). (Eilert et al., 2017) | http://dx.doi.org/10.17632/7ztzj63r68.1 |
| LabelLib | Modeling | LabelLib is a C++ library for the simulation of the accessible volume (AV) of small probes flexibly coupled to biomolecules.(Kalinin et al., 2012; Dimura et al., 2016) | https://github.com/Fluores-cence-Tools/LabelLib |
| FRETrest in Amber20 | Modeling | FRETrest is a set of helper scripts for generating FRET-restraints for Molecular Dynamics (MD) simulations performed with the AMBER Software Suite. (Dimura et al., 2019) | http://am-bermd.org/doc12/Amber20.pdf |

*6.6 Guidelines and recommendations of using smFRET for integrative structural biology*

One important example for developing recommendations and standards for the commu-nity comes from the Worldwide Protein Data Bank (wwPDB)(Berman, Henrick and Nakamura, 2003; Young et al., 2019). Given the importance of integrative structures for advancing life sci-ences and the significant worldwide investment made to determine them, the wwPDB initiated an effort to address the key challenges in enhancing its data-processing pipeline to accommodate integrative structures(Sali et al., 2015). Structural models and kinetic networks obtained using FRET experiments could be deposited in the prototype archiving system PDB-Dev for integrative or hybrid modeling (IHM)(Vallat et al., 2018). The IHM Task Force of the wwPDB has suggested to all methodologic communities, including FRET, that they should develop specific recommen-dations for their measurements, their analyses and how they are reported and documented fol-lowing the general guidelines summarized(Berman et al., 2019). In the long term, these recom-mendations could develop towards a standard for good scientific practice. The working principle of PDB-Dev relies on federating structural models and experimental data with appropriate data exchange, a proposal to establish a Fluorescence Biological Data Bank (FlBDB) for the archiving of data from fluorescence experiments is currently in progress.

As a starting point, the PDB-Dev and the Seidel group have recently developed a diction-ary for FRET data (https://github.com/ihmwg/FLR-dictionary) as a method-specific extension to the existing IHM dictionary so that FRET-restrained I/H structure models can now be deposited. To ensure that the reported results are reproducible, it is important that the raw data is sufficiently annotated. This includes reporting the sample details (e.g., preparation, purification and charac-terization, dye labeling positions and chemistry), instrumentation (e.g., setup components and acquisition settings), measurement conditions (e.g., buffer composition, temperature, ligands and other additives) and analysis (e.g., procedures and software used, information on data quality and the precision and accuracy of obtained results). In addition, correction factors, calibration param-eters and reference measurements should be specified and any assumptions that enter the anal-ysis should be provided. If structural modeling is performed with the obtained inter-dye distance information (mean distances or distance distributions(Haas *et al.*, 1975)), the corresponding procedures, tools, and the assessment of the uncertainty of the distances should be described to allow for validation of deposited structures.

## 7. Proposed actions to further establish the community and practice Open Science

To best achieve a consensus on the future directions of the smFRET community, an open forum is needed where the current issues, needs, and desires could be discussed. We propose the following tools to organize the community around standardization efforts and open science practices. Towards this end, the tools below have been proposed and some have already been put in place.

### 7.1 Community website as a central hub

A website for the FRET community has been established at https://www.fret.community. The community is open to everybody and registered members can populate their user profiles with additional information, such as a description of their scientific interests or a list of key publications. Besides providing regular updates on the activities within the community, the website also provides resources such as a curated list of software packages (see Table 1) and offers a discussion platform through an integrated forum. The website is now moderated by the advisory board of the community.

### 7.2 Listserv

To facilitate the dissemination of important information to the FRET community, an electronic mailing list (Listserv) has been established. In order to subscribe to it, smFRET practitioners are requested to register (free of charge) using the following link: https://www.fret.community/register. Through the email list, the members will be informed of ongoing activities and developments within the community, such as experimental or computational challenges, key publications in the fields, and workshops or meetings.

### 7.3 Server, website and repository

The website also serves as a platform for ongoing discussions, announcements of accepted relevant papers, announcements of upcoming meetings, workshops, competitions, joint publications etc. Moreover, a repository will be established, which will be accessible through the website, to host a collection of software packages and facilitate community-driven joint development of analysis tools. The repository will contain dedicated sections for acquisition software, raw data, analyses codes, analyzed data files, and file conversion utilities. In order to deposit code in the repository, documentation will be required (documentation guidelines will be provided).

The concept of the repository is to support open science and transparency. Anyone registered on the website will be able to access raw data, analyze and compare performances of the various analysis codes. Moreover, the codes could be updated and expanded (while keeping original versions) by anyone. This way, improvements and enhancements could be implemented and tested.

### 7.4 Participation in CASP and CASP-like competitions

Critical Assessment of protein Structure Prediction (CASP, http://predictioncenter.org/) is a grass-roots community-wide effort for predicting a three-dimensional protein structure from its amino acid sequence. CASP has been run, since 1994, as a double-blind competition. It provides research groups with an opportunity to objectively test their structure prediction methods. CASP has been exploring modeling methods based, in part, on sparse experimental data, including data from SAS, NMR, crosslinking, and FRET. This integrative CASP experiment was highlighted at the recent CASP13 meeting (www.predictioncenter.org/casp13), where the carbohydrate-binding module (CBM56) of a β-1,3-glucanase from Bacillus circulans with 184 amino acids (18.9 kDa) was studied as the first FRET data-assisted target F0964 in CASP13.

We propose that members of the smFRET community interested in using smFRET to study integrative structural biology participate in the CASP competition in several stages: (1) In the first stage, the smFRET community will only submit distances that will be evaluated with respect to the known (but undisclosed) crystal structure. The typical timeline of CASP is very challenging for experiment-based methods that involve the preparation of labeled samples, measurement, and analysis of FRET data. Moreover, experimental design works best when non-FRET-assisted structural predictions are already available. Those targets that are identified as difficult by the predictors and for which multiple possible folds are submitted without a clear winner, a FRET-assisted round could be insightful, as was the case for the first target CBM56. Moreover, we also suggest using the data from CBM56 and other FRET data sets for the CASP commons as joint training tools for the CASP community on FRET-assisted structural modelling; (2) In the second stage, the smFRET community will be full participants, using distance restraints + structural modeling (whether smFRET-assisted modeling or naïve modeling using smFRET information for validation) and will submit solved structures. These recommendations apply mostly to present and future practitioners of smFRET-driven integrative modelling. However, smFRET is just one out of many biophysical techniques that can provide experimental restraints in integrative modelling (XL-MS, single-particle cryo-EM, NMR, SAXS). Therefore, we propose that at a later stage, an all-biophysics integrative structural biology competition will be established.

### 7.5 smFRET meetings

Satellite meetings to the Conference on Methods and Applications in Fluorescence (MAF) have been organized to discuss practices, standards, competitions, and joint publications. We envision an occasional dedicated meeting for the smFRET community, such as the Bunsen meetings on FRET that have been held in 2011 and in 2016 at the Max Planck Institute for Biophysical Chemistry in Göttingen, Germany (link: http://fret.uni-duesseldorf.de/cms/home.html). However, to open these meetings to smFRET practitioners outside Europe, we propose to change the venue between continents. We also suggest using the satellite meetings and workshops to disseminate information (details of accurate FRET measurements, common practices, standards, and competitions) to young researchers and give them the chance to interact with the leading experts in the field.

The academic lifestyle in the post COVID19 pandemic era renders attendance in these upcoming smFRET meetings relatively difficult. A proper adaptation to the post COVID19 era might be in the form of smFRET webinars and web conferences that are open to all will provide FRET researchers the unique opportunity to listen and socialize virtually. These online seminars could also be a good practice of open science of the FRET community.

## 8. Summary

In this article, we have summarized our perspectives on the status of the smFRET field, limitations that still need to be overcome, and communal efforts of the smFRET community towards the adoption of consistent methodologies and open science practices. Such a community effort is necessary to solidify smFRET as an indispensable tool for dynamic structure determination. Based on the current needs and the actions that have already been taken, we have proposed additional actions for the near future. Our hope is that these efforts will benefit not only the smFRET community, but also the structural biology community and science overall.

## Author Contributions

All co-authors of this work have contributed to writing this manuscript.

## Acknowledgements

We wish to thank Niko Hildenbrandt and Sonja Schmid for fruitful discussions. R.B. wishes to thank Fabio D. Steffen for insightful discussions. This position paper was supported by the National Institutes of Health (NIH, grant R01 GM130942 to S.W., and to E.L. as a subaward; grant R01 GM095904 to X.M.; grants R01 GM079238 & R01 GM098859 to S.C.B.), the National Science Foundation (NSF, grants MCB-1818147 & MCB-1842951 to S.W.), the Human Frontier Science Program (HFSP, grant RGP0061/2019 to S.W.), the European Research Council (ERC; grant numbers 638536 & 860954 to T.C.; grant SMPFv2.0 to E.A.L.; grant No. 671208 (hybrid-FRET) to C.A.M.S.), the Deutsche Forschungsgemeinschaft (DFG, grants GRK2062 (C03) & SFB863 (A13) to T.C.; grants SPP2191 402723784 & SFB 1129 240245660 to E.A.L.; grant SE 1195/21-1 (SPP 2191) to C.A.M.S.; Germany's Excellence Strategy ,CIBSS – EXC-2189 – project ID 390939984 to T. Hugel), the Wellcome Trust (grant 110164/Z/15/Z to A.K.), the Swiss National Science Foundation (to B.S.), the UK Biotechnology and Biological Sciences Research Council (grant BB/S008896/1 to A.K.; grant BB/T008032/1 to T.D.C.), the Royal Society (grant RGS\R2\180405 to T.D.C.), the Agence Nationale de la Recherche (grants ANR-17-CE09-0026-02, ANR-18-CE11-0004-02 & ANR-19-CE44-0009-02 to E.M.), the Israel Science Foundation (ISF, grant 1250/19 to G.H.), the National Research Foundation of Korea (grant 2019R1A2C2090896 to N.K.L.), the Independent Fund Denmark (grant 6110-00623B to V.B.), the Milner Fund (to E.L.), the Hebrew University of Jerusalem (start-up funds to E.L.), the Ludwig Maximillian Universität (LMU, the LMUinnovativ program BioImaging Network, BIN, to D.C.L.), the Biological Optical Microscopy Platform (BOMP) at University of Melbourne (to H.S.), the University of Applied Sciences Mittweida and the Alfred-Werner-Legat of the Department of Chemistry, University Zurich (financial support to R.B.), the EPSRC DTP Studentship (to B.A.) and the CAS[LMU] fellowship (to T.C., S.W., E.L.). G.H. holds the Hilda Pomeraniec Memorial Professorial Chair. This work was performed under the auspices of the U.S. Department of Energy by Lawrence Livermore National Laboratory under Contract DE-AC52-07NA27344 (for T.L.).

## Competing interests
The authors declare that no competing interests exist.